\documentclass[aps,longbibliography,preprint,showkeys]{revtex4-1}

\usepackage{graphicx}

\usepackage{algorithm2e}
\RestyleAlgo{ruled}
\usepackage{algorithmic}

\usepackage{amsmath,amsfonts,amssymb}
\usepackage{bm}
\usepackage{booktabs}
\usepackage[ plainpages = false, pdfpagelabels,
	     bookmarks,
	     bookmarksopen = true,
	     bookmarksnumbered = true,
	     breaklinks = true,
	     linktocpage,
	     colorlinks = false,  
         hypertexnames=false,   
	     linkcolor = blue,
	     urlcolor  = blue,
	     citecolor = red,
	     anchorcolor = green,
	     hyperindex = true,
	     hyperfigures
	     ]{hyperref}
\usepackage{tikz}
\usepackage[capitalise,noabbrev]{cleveref}

\usepackage{comment}

\usepackage{soul}
\usepackage{stmaryrd}
\usepackage{subcaption}
\usepackage{textcomp}
\usepackage{xcolor}
\usepackage{natbib}
\usepackage{pgfplots}
\pgfplotsset{compat=newest}
\usetikzlibrary{plotmarks}
\usetikzlibrary{arrows.meta}
\usepgfplotslibrary{patchplots}
\usepackage{grffile}
\usepackage{amsmath}
\pgfplotsset{plot coordinates/math parser=false}
\newlength\figureheight
\newlength\figurewidth
\pgfplotsset{every axis/.append style={
                    label style={font=\small},
                    tick label style={font=\footnotesize},
                    legend style={font=\tiny},
                    title style={font=\small}
                    }}
\usetikzlibrary{plotmarks}
\pgfplotsset{minor grid style={dotted,gray}} 
\pgfplotsset{every axis/.append style={thick, tick style=semithick}}
\pgfplotsset{every mark/.append style={mark size=1pt}}
\usepgfplotslibrary{groupplots}
\pgfplotsset{xticklabel style={/pgf/number format/fixed,
        /pgf/number format/precision=5},yticklabel style={/pgf/number format/fixed,
        /pgf/number format/precision=5}}

\usepackage{babel,rotating,dcolumn}
\usepackage{colordvi,graphicx,color,amsbsy,amsmath,bm,amssymb}
\usepackage{type1cm}

\usepackage{amsthm}

\theoremstyle{remark}
\newtheorem{remark}{Remark}

\renewcommand{\d}{\, \mathrm{d}}                                   
\newcommand{\vect}[1]{\textrm{\boldmath${#1}$}}                 

\newcommand{\scalar}{f}
\newcommand{\domain}{\Omega}
\newcommand{\defeq}{:=}

\newcommand{\map}{\vect{X}}

\newcommand{\vmap}{\map}

\newcommand{\error}{\varepsilon}

\newcommand{\Kn}{\mathrm{Kn}}

\newcommand{\ekin}{e_\mathrm{kin}}
\newcommand{\energy}{\mathcal{E}}

\newcommand{\Etot}{\energy}                 
\newcommand{\Epot}{\energy_\mathrm{pot}}                 
\newcommand{\Ekin}{{\energy}_\mathrm{kin}}               
\renewcommand{\u}{\vect{u}}                              

\newcommand{\RomanNumeralCaps}[1]                               
    {\MakeUppercase{\romannumeral #1}}
\newcommand{\define}[1]{\textit{#1}}                            

\begin{document}

\title{Characteristic Mapping Method for Vlasov--Poisson\\ with BGK-collisions}

\author{Xi-Yuan Yin$^{1}$, Philipp Krah$^{2}$, Zetao Lin$^{3}$, Jean-Christophe Nave$^{4}$ and Kai Schneider$^{3}$}

\affiliation{$^{1}$ Max Planck Institute for Mathematics in the Sciences,
Inselstrasse 22, 04103 Leipzig, Germany}

\affiliation{$^{2}$ IRFM - CEA Cadarache, 13108 Saint-Paul-lez-Durance, France}

\affiliation{$^{3}$ Institut de Math\'ematiques de Marseille (I2M), Aix-Marseille Universit\'e, CNRS, 3 place Victor Hugo, 13331 Marseille Cedex 3, France}

\affiliation{$^{4}$ McGill University, Department of Mathematics and Statistics, 805 Sherbrooke Street West, Montreal, Quebec H3A 0B9, Canada \email{jcnave@math.mcgill.ca}}

\date{\today}


\begin{abstract}
  This work presents the first steps for simulating kinetic plasmas with collisions using the characteristic mapping method (CMM). The CMM is a semi-Lagrangian method that uses the group structure of volume-preserving diffeomorphisms to efficiently represent the flow-maps of transport problems. Individual submaps can be composed to follow the flow backward in time to its initial foot point, numerically establishing the back-to-label symmetry. The novelty of the presented work is the handling of the source term by storing the Duhamel integrals of each time sub-interval, allowing efficient temporal-integration. Furthermore, we use the Lagrangian structure to avoid implicit time-stepping in the hydrodynamic regime, which is known to be stiff. We benchmark our method on the Boltzmann--BGK and Vlasov--Poisson--BGK equations and consider different test cases, the Sod shock tube problem and nonlinear Landau damping for different Knudsen numbers. We show third-order spatial and temporal convergence and illustrate the fine-scale zoom property of CMM for the bump-on-tail instability.
\end{abstract}


\maketitle


\newpage
\section{Introduction}

The general motivation of this paper are kinetic equations with source terms, which have numerous applications in science and engineering. They are challenging for numerical simulations due to their computational complexity. Their solution, i.e., the particle distribution function, satisfies a transport equation and is a function depending on several variables in phase space. Problems are thus typically high-dimensional, up to six dimensions in phase space. 
Taking collisions into account results in source terms which make the equations stiff, in particular when the Knudsen number (a dimensionless parameter given by the ratio between the molecular mean free path and characteristic length scales) tends to zero and the system enters the regime of continuum mechanics.
Modeling of collisions is of huge interest in magnetically confined fusion, astrophysics and for reentry problems of spacecraft.
There are different self-consistent models, (e.g., for rarefied gas dynamics and plasma physics) and many collision operators are available, going in decreasing complexity from nonlinear Boltzmann collision operators, which is the full binary-collision integral with different collision kernels (e.g. for hard spheres, etc.), via linear Boltzmann operators to relaxation models toward equilibrium. For details, we refer to the book by Cercignani \cite{cercignani1988boltzmann}.
In the present work, we consider the simple Bhatnagar–Gross–Krook (BGK) \cite{BhatnagarGrossKrook1954} collision term, a local relaxation model towards the equilibrium Maxwellian distribution.

\medskip
Multiple numerical challenges are encountered, e.g., the generation of very fine scales due to plasma filamentation, respecting physical bounds of the solution and conservation properties.
To this end, numerous numerical approaches have been developed for kinetic equations with source terms, including Eulerian and semi--Lagrangian methods, Monte-Carlo methods (DSMC), discrete velocity methods to mention the most important ones. Often, time splitting is used and asymptotic preserving schemes are applied.
A complete overview is beyond the scope of the paper and for reviews, we refer for instance to \cite{mouhot2006fast, kowalczyk2008numerical, rodriguez2011overview, aristov2001direct, dimarco2014numerical}.
An overview on Monte--Carlo methods for Boltzmann can be found, e.g. in \cite{pareschi2001introduction}. They have the advantage that they work in high dimensions and can be parallelized. However, they have low accuracy and they suffer from numerical noise. Wavelet-based techniques have been developed to improve their convergence, see, e.g., \cite{nguyen2010wavelet}.
Motivated by their high accuracy, different spectral methods have been designed, for instance, in
\cite{gamba2017fast} for the Boltzmann equation with integral collision operators for arbitrary collision kernels,
in \cite{gamba2009spectral} for hard sphere collision kernels,
in \cite{hu2020fast} for non-cutoff kernels with non-integrable singularities and
in \cite{pareschi2000numerical} a spectral scheme for the evaluation of the collisional integral.
A discontinuous Galerkin method has been likewise proposed \cite{ding2021semi} using a stiffly accurate diagonally implicit Runge--Kutta (DIRK) method along characteristics for time integration. The latter is also used in the present work in combination with the characteristic mapping method (CMM).
%
%
CMM is a semi--Lagrangian method used to solve transport equations with high precision that was first introduced for advection equations, \cite{mercier2020characteristic}. It is based on the gradient--augmented level set method \cite{nave2010gradient,seibold2011jet,chidyagwai2011comparative} and remapping techniques from \cite{kohno2013new}. The original CMM approach was later extended to nonlinear transport equations, in particular for two- and three-dimensional incompressible Euler in \cite{YinMercierYadavSchneiderNave2021} and \cite{yin2023characteristic}, respectively.
A generalization for studying the hydrodynamics on the rotating sphere has been presented in \cite{taylor2023characteristic}.
A crucial feature of the CMM is the possibility of compositional adaptivity, i.e., the backwards flow map can be decomposed into submaps. As a result, the method possesses exponential resolution in linear time. Composing the submaps, each of which is represented by Hermite polynomials, then yields a high-order approximation and the resulting composed map may be evaluated on any arbitrary grid. This property allows us to zoom into the fine-scale structure of the solution by evaluating the initial condition on arbitrary grids, which is assumed to be known analytically.

More recently, kinetic equations have also been studied using the CMM, for instance, the Vlasov--Poisson equation in \cite{KrahYinNaveBergmannSchneider2024}.
In \cite{krah2026hybrid} Krah et al. have introduced a hybrid approach for kinetic equations by coupling the CMM with the numerical flow iteration proposed in \cite{kirchhart2024numerical}.
Further extensions of CMM concern flows on manifolds with applications to flows on the rotating sphere \cite{taylor2023characteristic}, 
and also in the context of machine learning for diffeomorphic neural operators \cite{taylor2025diffeomorphic}. 

A method for integrating source terms using the Duhamel principle was introduced by Yin in \cite{yin2021thesis} and was further developed for ideal magnetohydrodynamics in \cite{YinKrahNaveSchneider2026}.
The main objective of this paper is to present the extension of the CMM for transport equations with source terms for kinetic equations with collisions. One numerical difficulty arises in the limit of small Knudsen numbers when the relaxation time scale is much shorter than the advection time scale, making the time-stepping stiff. The CMM circumvents this stiffness by following Lagrangian trajectories, effectively decoupling the advection from the collisions. We also propose time-stepping schemes using diagonally implicit Runge--Kutta and using exponential integrators to increase the accuracy of the source term integration. The handling of Dirichlet boundary conditions with the CMM, necessary to impose in- and out-flow conditions, is also studied. A Fourier pseudo-spectral method using a domain periodization trick with a smooth periodic continuation of the velocity (see ref.~\cite{KrahYinNaveBergmannSchneider2024}) was also implemented as a benchmarking tool for test cases with sufficiently large Knudsen numbers.


The remainder of the paper is organized as follows. In sec.~\ref{sec:kinetic} we present the governing kinetic equations, either Boltzmann or Vlasov--Poisson with the BGK collision operator. Section~\ref{sec:CMM} recalls the characteristic mapping method for kinetic equations first without source terms and then shows its extension that includes source terms using the Duhamel integral. The handling of the stiffness induced by the source term is also described. The numerical results for the classical Sod shock tube problem in 1D, nonlinear Landau damping in 1D+1D phase space and the bump--on--tail instability test cases are detailed in sec.~\ref{sec:pb}.
Finally, conclusions are drawn in sec.~\ref{sec:conclusion} and some perspectives are given for future studies.
The appendix summarizes technical parts on the reference pseudo-spectral method.

\section{Kinetic equations with source terms}
\label{sec:kinetic}

In this paper, we study the application of the CMM to kinetic equations with source terms, in particular with the BGK collision term. This will be studied in the single-species Boltzmann and Vlasov--Poisson cases in (1+1) dimensional phase space, let $\Omega_x = S^1$, the 1d circle represented as a periodic interval $[0, L]$, be the physical domain and $\Omega_v = \mathbb{R}$ be the velocity domain. In either equations, we denote by $f (x, v, t)$ the particle distribution function on $\Omega_x \times \Omega_v \times \mathbb{R}_+$.

The BGK forcing term 
\begin{equation}
    s[f] \defeq \frac{1}{\tau}(M[f] - f)
\end{equation}
describes the relaxation toward the Maxwell distribution $M[f]$ at a time scale $\tau$, the Knudsen number. The Maxwellian distribution $M[f]$ is determined at each point $x \in \Omega_x$ by the macroscopic state at that point:
\begin{equation}
    M[f](x, v)  \defeq \frac{\rho(x)}{\sqrt{2\pi R T(x)}} e^{-\frac{(v-u(x))^2}{2RT(x)}}\,,
\end{equation}
where $\rho$ is the macroscopic density, $u$ the macroscopic velocity, $T$ the macroscopic temperature and $R \approx 8.31446261815324$, the universal gas constant. These are given by the zeroth, first and second macroscopic moments of the distribution:
\begin{equation}
\label{eq:macromoments}
(\rho,\rho u, \rho \ekin ) \defeq \int_{\Omega_v}  f(x, v, t) \psi (v) \d v \quad \text{with}\quad \psi(v) = (1,v,v^2/2) ,
\end{equation}
and $T = \frac{2}{R}(\ekin - \frac12 u^2)$. We note that the operator $M$ does not change the macroscopic state and therefore $M[M[f]] = M[f]$.

\subsection{Boltzmann-BGK}
\label{sec:Bolzmann-BGK}

The Boltzmann-BGK equation for monoatomic gases in one dimension is given by \cite{BhatnagarGrossKrook1954}
\begin{equation}
    \partial_t f + v \partial_x f =s[f] .
\end{equation}
On the phase space, this equation describes the transport of particles according to their microscopic velocity $v$ while the distribution relaxes to the Maxwellian equilibrium at rate $\tau^{-1}$. The parameter $\tau$, which controls this relaxation rate, is called the Knudsen number, $\tau = \Kn$. We note that the characteristic curves of this advection are given by
\begin{equation}
 (x(t), v(t)) = (x_0 + v_0 t, v_0) \qquad x_0 = x(0), \, v_0 = v(0)   ,
\end{equation}
which corresponds to a purely horizontal shear transport on the phase space.

In the limit when the Knudsen number tends to zero, we have formally $f = M[f]$ at all points ($x, t)$, one obtains the compressible Euler equations of classical gas dynamics:
\begin{subequations}
\begin{align}
    & \partial_t \rho + \partial_x (\rho u) = 0 \\
    & \partial_t (\rho u) + \partial_x (\rho u^2 +p) = 0 \\
    & \partial_t \ekin + \partial_x (u (\ekin + p)) = 0\,,
\end{align}
\end{subequations}
the equations of conservation of mass, momentum and energy, respectively.

\subsection{Vlasov--Poisson--BGK}
\label{sec:VP-BGK}
The Vlasov--Poisson equation with BGK collisions couples the $x$ and $v$ dependence in phase space through the electric field:
\begin{equation}
\begin{aligned}
    & \partial_t f + v \partial_x f + \frac{q}{m} E \partial_v f =s[f] ,\\
    & \lambda^2\partial_{xx}\phi =  1 - \int_{\Omega_v} f \d v\, = 1 - \rho(x),
\end{aligned}
\end{equation}
where $f =f(x,v,t)$ is the particle distribution function, and $E(x,t) = -\partial_x \phi(x,t)$ is the electric field from an electric potential $\phi(x,t)$. Here, $m$ is the parameter for particle mass, $q$ for elementary charge and $\lambda$ for the normalized Debye length. In the rest of this paper, we set $m = q = \lambda = 1$ (see also ref.~\cite{LiuShiWanHeCao2020}). We will also denote the phase space velocity field $\u$, given by $\u[f] = (v, E)$. The equation in phase space is rewritten as
\begin{equation}
\label{eqn:VP_BGK}
\begin{aligned}
    & (\partial_t + \u[f] \cdot \nabla ) f = s[f] ,\\
    & \u [f] = (v, E[f]) \quad \text{where} \quad E[f] = \partial_x \Delta_x^{-1} (\rho(x) - 1) .
\end{aligned}
\end{equation}

We will be interested in the conservation of total energy in this system and in our numerical results. Consider the kinetic and potential energies $\Ekin$ and $\Epot$:
\begin{align}
   & \Ekin (t) = \int_{\Omega_x} \rho \ekin \d x = \int_{\Omega_x} \int_{\Omega_v} \frac{|v|^2}{2} f(x, v, t) \d v \d x , \\
   & \Epot (t) = \int_{\Omega_x} \frac{|E |^2}{2} \d x = - \int_{\Omega_x} \frac{\phi}{2} \partial_{xx} \phi \d x = \int_{\Omega_x} \int_{\Omega_v} \frac{\phi(x,t)}{2} f(x, v, t) \d v \d x .
\end{align}
Testing equation \eqref{eqn:VP_BGK} against $|v|^2 + \phi$, one sees that the total energy $\Etot = \Ekin + \Epot$ is formally a conserved quantity of the system. In the hydrodynamic limit of $\tau \to 0$, the limiting equations is the Euler-Poisson system.

\section{Characteristic mapping method for kinetic equations}
\label{sec:CMM}
The \define{characteristic mapping method} (CMM) is a semi--Lagrangian method for advection problems.
In this work, we consider generally an advection equation with source term
\begin{equation}
\label{eq-def:CMM-advect}
    \begin{aligned}
    & (\partial_t + \u[f] \cdot \nabla) f = s[f] \\
    & f(\vect{\xi},0) = f_0(\vect{\xi})           .
    \end{aligned}
\end{equation}
From now on, we denote by $\Omega = \Omega_x \times \Omega_v$ the phase space domain and $\vect{\xi} \in \Omega$ the phase space positions. The field $\u : \Omega \to \mathbb{R}^2$ is the phase space velocity which may depend on the particle distribution $f$ function on $\Omega$.

We see that in both the Boltzmann and Vlasov--Poisson cases, $\u$ is a divergence-free field and under suitable assumptions on the initial conditions and $\tau$, that $\u$ is sufficiently regular so that characteristic trajectories form a diffeomorphic flow-map which guarantees a relabeling symmetry between Eulerian and Lagrangian frames. In the following, we first recall the CMM for advection without source terms and with general source terms, and then describe the specific implementation used in the BGK collision term, which is \emph{a priori} stiff.

\subsection{CMM without source term}
\label{ssec:CMM-no-ST}
\begin{figure}[h]
  \centering
  \input{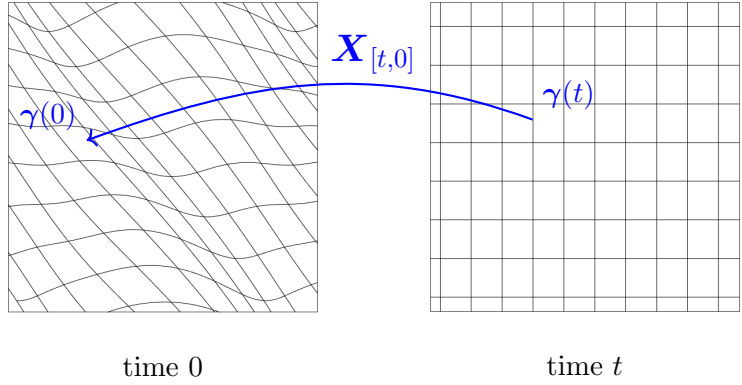}
    \caption{Illustration of the backward flow map.}
    \label{fig:illu_map}
\end{figure}

The concept of CMM is based on the methods of characteristics. 
In this method, one follows the trajectories, i.e., \define{characteristic curves} $\vect{\gamma}\colon\mathbb{R}_+\to \domain$ that satisfy the ODE:
\begin{align}
        \label{eq-def:characteristic_curve}
        \frac{\d \vect{\gamma}}{\d t}\, = {\u}   \quad \text{with} \quad
        \vect{\gamma}(0) = \vect{\xi}\in\domain\,.
\end{align}
Considering the absence of a source term ($s=0$) we obtain:
\begin{equation}
    \label{eq:ODE-characterisitcs_without_S}
    \frac{\d {\scalar}}{\d t}(\vect{\gamma}(t),t) = \partial_t\scalar + \frac{\d \vect{\gamma}}{\d t}\cdot\nabla\scalar = 0\,.
\end{equation}
This means that along the characteristic curves, the solution stays constant in time:
\begin{equation}
\label{eq:ccurves}
    \scalar(\vect{\gamma}(t),t)=\scalar_0(\vect{\xi})\,.
\end{equation}
Evolving the solution of \eqref{eq-def:characteristic_curve} for all $\vect{\xi}\in\domain$ simultaneously is done using the \define{characteristic map} $\vect{\map}\colon\domain \times \mathbb{R}_+\to \domain$, defined as 
\begin{equation}    
\label{eq-def:CMM-map}
\map(\vect{\gamma}(t),t)=\vect{\gamma}(0)=\vect{\xi}  \qquad \text{for any}\quad\vect{\xi}\in \domain \, .
\end{equation}
 Therefore, the solution of \eqref{eq-def:CMM-advect} at time $t$ can be computed by the composition with the initial condition
\begin{equation}
    \label{eq:cmap}
    \scalar(\vect{\xi},t)=\scalar_0(\map(\vect{\xi},t))\,.
\end{equation}
Note that in contrast to equation \eqref{eq:ccurves}, \eqref{eq:cmap} follows the characteristic backward in time.
Therefore, $\map$ is often called 
\define{backward map} (see Fig.~\ref{fig:illu_map}).

We can more generally define the backward map $\map_{[t, s]}$ in any arbitrary time interval $[s, t]$, by requiring that for all characteristic curves $\vect{\gamma}$, we have $\map_{[t, s]}(\vect{\gamma}(t)) = \vect{\gamma}(s)$. We invert the positions of $s$ and $t$ in $\map_{[t, s]}$ as reminder that it is a backward map. This family of diffeomorphisms then satisfies the equations
\begin{subequations}\label{eq-def:CMM-map-advect}
    \begin{align}
        & (\partial_t + \u \cdot \nabla) \map_{[t, s]} = 0 ,\\
        & \partial_s \map_{[t, s]} = \u( \map_{[t, s]}, s ) \\
        & \map_{[s, s]}(\vect{\xi}) = \map_{[t, t]} (\vect{\xi}) = \vect{\xi} .
    \end{align}
\end{subequations}

\subsubsection{Group property and remapping}
\label{sssec:semi-group-no-source}

As introduced in \cite{YinMercierYadavSchneiderNave2021}, the CMM exploits the group property of the characteristic map, which allows the decomposition of the flow map using a decomposition of the time interval $[0, t]$ into subintervals $0 = T_0 < T_1 < T_2 < \cdots, < T_{m} \leq t$. Denote by $\map_{[t, s]}$ the backward map from time $t$ back to time $s$, we have

\begin{equation}
\vmap_{[t, 0]} = \vmap_{[T_1, 0]}\circ\vmap_{[T_2,T_1]}\circ \dots \circ \vmap_{[T_{m},T_{m-1}]}\circ\vmap_{[t, T_{m}]}
\label{eq:semi-group}
\end{equation}
where each \define{submap} solves:
\begin{equation}
    \begin{aligned}
        & (\partial_t  + {\u} \cdot \nabla ) \vmap_{[t, T_{i-1}]} = 0 \quad \text{for } (\vect{\xi},t)\in\domain\times[T_{i-1},T_i] \\
        &\vmap_{[T_{i-1}, T_{i-1}]}(\vect{\xi}) =\vect{\xi}\,.
    \end{aligned}
    \label{eq:CMM-submap}
\end{equation}

This decomposition controls the error $\error$ that occurs during the numerical integration of \eqref{eq-def:CMM-map-advect}, by remapping, i.e., limiting the solution time of each submap by reinitialization.
Thus, starting from the time $t=T_{i-1}$ the submap is initialized as the identity map and \eqref{eq:CMM-submap} is integrated in time until a certain error threshold is reached.

For the CMM without source terms, we typically use the \define{volume-conservation error} as remapping criterion, requiring that
\begin{equation}
    \error_\text{det}\defeq \Vert \mathrm{det} (\nabla \vmap_{[T_{i-1},T_{i}]}) - 1 \Vert_\infty\le \delta_\text{det}\,,
\end{equation}
limited by the \define{incompressibility error threshold} $\delta_\text{det}$.

\subsection{CMM with source term}
\label{ssec:CMM-ST}
In a recent publication \cite{YinKrahNaveSchneider2026} CMM was extended to source terms in the context of ideal MHD. Here we briefly explain the concept and its adaptation towards kinetic equations.

When including a source term, \eqref{eq:ODE-characterisitcs_without_S} needs to take into account the change of the function along the characteristics:
\begin{equation}
    \label{eq:ODE-characterisitcs_with_S}
    \frac{\d {f}}{\d t}(\vect{\gamma}(t),t) = \partial_t f + \frac{\d \vect{\gamma}}{\d t}\cdot\nabla f = s[f]\,
\end{equation}
and thus the solution is no longer constant along the characteristics:
\begin{equation}
\label{eq:ccurves-s}
    f (\vect{\gamma}(t),t)= f_0(\vect{\gamma}_0)+\int_0^ts[f(\cdot,r)](\vect{\gamma}(r)) \d r \,.
\end{equation}

Using the backward map to keep track of all backward in time trajectories, the solution can be expressed as
\begin{equation}
\label{eq:ccurves-s-m}
    f (\vect{\xi},t) = f_0(\map_{[t, 0]}(\vect{\xi}))+\int_0^t s[f(\cdot , r)] \circ \map_{[t, r]} \d r \,,
\end{equation}
where the source term time-integration essentially arise from Duhamel's principle using right-composition by $\map$ as the solution operator to the homogeneous problem.

\subsubsection{Source term integration and decomposition}
\label{sssec:semi-group-source}

A similar decomposition can be applied to the source term integration. Indeed, as the time integral in \eqref{eq:ccurves-s-m} essentially solves a forced advection equation with zero as initial data, a discretization on a fixed grid may eventually run out of spatial resolution and deteriorate the arbitrary resolution provided by the CMM. Employing the submaps from the decomposition above, we can rewrite the source term integral as
\begin{equation}
    \int_0^t s[f(\cdot , r)] \circ \map_{[t, r]} \d r = \sum_{i=1}^m \left( \int_{T_{i-1}}^{T_i} s[f(\cdot , r)] \circ \map_{[T_i, r]} \d r \right) \circ \map_{[t, T_i]} +  \int_{T_{m}}^{t} s[f(\cdot , r)] \circ \map_{[t, r]} \d r.
\end{equation}

Keeping the same notation as for the flow-map, we denote by $F_{[t_1, t_0]}$ the subinterval source term integral
\begin{equation}
    F_{[t_1, t_0]} \defeq \int_{t_0}^{t_1} s[f(\cdot , r)] \circ \map_{[t_1, r]} \d r
\end{equation}

Then, for the numerical scheme, we solve in each subinterval the following system of equations
\begin{equation}
    \begin{aligned}
        & (\partial_t  + {\u} \cdot \nabla ) \vmap_{[t, T_{i-1}]} = 0 \quad & \text{for } (\vect{\xi},t)\in\domain\times[T_{i-1},T_i] \\
        & (\partial_t  + {\u} \cdot \nabla ) F_{[t, T_{i-1}]} = s[f] \quad & \text{for } (\vect{\xi},t)\in\domain\times[T_{i-1},T_i] \\
        &\vmap_{[T_{i-1}, T_{i-1}]}(\vect{\xi}) =\vect{\xi} , & \\
        &F_{[T_{i-1}, T_{i-1}]}(\vect{\xi}) = 0 . & \\
    \end{aligned}
    \label{eq:CMM-submap-source}
\end{equation}
with
\begin{equation} \label{eqn:recursiveCompositionEvaluation}
\begin{aligned}
    f(\vect{\xi}, t) & = f_0 \circ \map_{[t, 0]} + \sum_{i=1}^{m-1} F_{[T_i, T_{i-1}]} \circ \map_{[t, T_i]} + F_{[t, T_{m-1}]} \\
    & = \left( \left( \cdots \left( \left( f_0 \circ \map_{[T_1, 0]} + F_{[T_1, 0]}  \right) \circ \map_{[T_2, T_1]} + F_{[T_2, T_1]} \right) \cdots \right) \right. \\ 
    &\left. \qquad \circ \map_{[T_{m-1}, T_{m-2}]} + F_{[T_{m-1}, T_{m-2}]} \right) \circ \map_{[t, T_{m-1}]} + F_{[t, T_{m-1}]} ,
\end{aligned}
\end{equation}
where the second line writes the sum as a recursive composition-accumulation operation used in the numerical implementation.

We note that since the initial condition for each source term subintegral $F_{[T_i, T_{i-1}]}$ is 0, the spatial oscillations of this term can be controlled by the length of the time subinterval and therefore appropriate spatial resolution for the evolution of these terms can be guaranteed using sufficiently frequent remappings.

\subsubsection{Exponential decay from the BGK source term}
\label{sssec:exp-decay-source}

The BGK source term source term studied in this paper, encodes an exponential decay of the history of the distribution, which gets replaced by the Maxwellian at the current time. The time-scale of this relaxation is given by the Knudsen number $\tau$ which is independent from the advection time scale, this typically leads to numerical stiffness. In the context of CMM, the advection is computed by following characteristic trajectories meaning that the transport and decay are decoupled at the discretization level. Indeed, writing the equation as
\begin{equation}
    ( (\partial_t + \frac{1}{\tau}) + \u [f] \cdot \nabla ) f = \frac{1}{\tau} M [f]
\end{equation}
we see that the natural \emph{pullback/solution} operator to the decaying advection should be $f_0 \mapsto \exp(-t/\tau) f_0 \circ \map_{[t, 0]}$ rather than the typical $f_0 \mapsto f_0 \circ \map_{[t, 0]}$. This can also be understood in terms of multiplying by an exponential integrating factor, similar to the exponential integrator schemes \cite{hochbruck2005exponential}.

We can reformulate the exponential decay in the pullback operator as follows: Along a characteristic curve $\vect{\gamma}$, we have that $f(\vect{\gamma}(t), t)$ satisfies the ODE
\begin{equation}
    f' + \frac{1}{\tau} f - \frac{1}{\tau} M[f] = 0
\end{equation}
as 1st order ODE where we can use the integrating factor $\exp(\frac{t}{\tau})$. Therefore, let $g(\bm{\xi}, t) = \exp(\frac{t}{\tau}) f(\bm{\xi}, t)$, we have that
\begin{equation}
    (\partial_t + \u \cdot \nabla) g = \exp(t / \tau)(\partial_t + \u \cdot \nabla) f + \exp(t / \tau)\frac{1}{\tau}f = \exp(t / \tau)\frac{1}{\tau} M .
\end{equation}
So it is sufficient to apply the CMM with source term to the system
\begin{equation}
    \begin{aligned}
        & (\partial_t + \u \cdot \nabla) g = \frac{\exp(t/\tau)}{\tau} M [ f ] ,\\
        & f = \exp(-t/ \tau) g , \quad  g(\cdot, 0) = f(\cdot, 0).
    \end{aligned}
\end{equation}

This shows that the advection and decay from the source term can be decoupled by providing an appropriate approximation for the velocity field $\u = \u [f]$, this is a similar approach to the one used in \cite{YinMercierYadavSchneiderNave2021, yin2023characteristic} to linearize the Euler equations around each time $t$ by a time-extrapolation of the velocity from its history. In the case of the Vlasov--Poisson--BGK equations, we will provide this time-extrapolation using successive time-stepping 
with a lower order method. This is in part because we expect that a simple polynomial extrapolation might not work well when the density exhibits exponential decay in time.

Rewriting the equation with this integrating factor, we arrive to the following time-evolution equation for the CMM with source terms:
\begin{equation} \label{eqn:expTimeStepping}
    \begin{aligned}
        & (\partial_t  + \u [ f ] \cdot \nabla ) \map_{[t, 0]} = 0 , \\
        & f(t) = \exp (- t/\tau ) f_0 \circ \map_{[t, 0]} + \int_0^t \exp (- (t-s)/\tau ) M[ f(s) ] \circ \map_{[t, s]} ds .
    \end{aligned}
\end{equation}
Applying a time decomposition $0 = T_0 < T_1 < \cdots < T_m < t$ and letting
\begin{equation} 
    F_{[t_1, t_0]} =   \int_{t_0}^{t_1}  \exp (- (t_1-s)/\tau ) M[ f(s) ] \circ \map_{[t, s]} ds .
\end{equation}
we have
\begin{equation} \label{eqn:decomposedDistEval}
    f(t) = F_{[t, T_m]}  +  \sum_{i=1}^m \exp(-(t-T_i)/\tau) F_{[T_i, T_{i-1}]} \circ \map_{[t, T_i]} + \exp (- t/\tau ) f_0 \circ \map_{[t, 0]},
\end{equation}
and the following evolution equation for the source term accumulation $F_{[t, T_m]}$. Let $G_{[t, T_m]} = \exp \left( \frac{t-T_m}{\tau} \right) F_{[t, T_m]}$, we have
\begin{equation}
    (\partial_t  + \u [ f ] \cdot \nabla )  G_{[t, T_m]}  = \exp \left( \frac{t-T_m}{\tau} \right) M [f] .
\end{equation}

This in principle removes the stiffness from the ODE integration of $F$, the exponential only appearing in the source term. In this case the numerical difficulty should lie in the accurate integration of the exponential using quadrature rather than a stability problem.


\subsection{Source term integration using exponential coordinates}
\label{ssec:CMM_ST_exp}

This section describes the time-stepping scheme used to compute the one-step solution $(\map_{[t^{n+1}, t^n]}, F_{[t^{n+1}, t^n]})$ given an approximate solution
\begin{equation} \label{eqns:approx_onestepsoln}
    \begin{aligned}
        & \tilde{f} (t) & \text{for } \: t \in [t^n, t^{n+1}], \\
        & \tilde{\map}_{[t^{n+1}, t]}  & \text{for } \: t \in [t^n, t^{n+1}].
    \end{aligned}
\end{equation}

The approximate solution allows for a linearized evaluation of the following integrals:
\begin{equation} \label{eqns:linearized_timestep}
    \begin{aligned}
        & \map_{[t^{n+1}, t^n]}(\bm{\xi})  \approx \bm{\xi} - \int_{t^n}^{t^{n+1}} \tilde{\u} ( \tilde{\map}_{[t^{n+1}, t]}(\bm{\xi}), t) dt , \\
        & G^{n+1} \approx G^n \circ \map_{[t^{n+1}, t^n]} + \int_{t^n}^{t^{n+1}} \frac{\exp( t/ \tau)}{\tau} M [ \tilde{f} (t) ] \circ \tilde{\map}_{[t^{n+1}, t]} dt .
    \end{aligned}
\end{equation}

We note that the integrand in the second line grows exponentially so that uniform quadrature will not be accurate for small $\tau$. We therefore apply a substitution
\begin{equation*}
     s = \exp((t-t^n)/ \tau), \quad ds = \tau^{-1} \exp((t-t^n)/ \tau)  dt, \quad t(s) = t^n + \tau \log(s) 
\end{equation*}
so that
\begin{equation} \label{eqn:onestep_integral}
    \int_{t^n}^{t^{n+1}} \frac{\exp( t/ \tau)}{\tau} M [ \tilde{f} (t) ] \circ \tilde{\map}_{[t^{n+1}, t]} dt = \exp(t^n/\tau) \int_{1}^{\exp(\Delta t/ \tau)} M [ \tilde{f} (t(s)) ] \circ \tilde{\map}_{[t^{n+1}, t(s)]} ds,
\end{equation}
to which a standard quadrature rules may be applied. This change of variable in time is similar to the approach of exponential Runge--Kutta methods \cite{ando2026exploring, hochbruck2005exponential}. A similar approach using a change of variable in time also appears in fluid simulations with obstacle penalization \cite{schneider2005numerical} where a still source term is introduced and where adaptive time-stepping is introduced.

We note that if $\tilde{f}$ is obtained from an order $\alpha$ scheme from $f^n$, the error on the approximations $\tilde{\u}, \tilde{f}$ and $\tilde{\map}$ are given by the local truncation error, i.e., $\mathcal{O} ( \Delta t^{\alpha + 1})$. This implies that if the above integrals are then computed using an order $\alpha +1$ quadrature, the local truncation error formally becomes $\mathcal{O} ( \Delta t^{\alpha + 2})$, yielding an order $\alpha+1$ scheme overall.

From, \eqref{eqn:onestep_integral}, given quadrature points $t_i \in [t^n, t^{n+1}]$ with associated $a_i^l$ and $a_i^e$ the quadrature weights for the linear and exponential coordinates, the time update scheme has the following form:
\begin{equation} \label{eqns:timestepScheme}
    \begin{aligned}
        & \map_{[t^{n+1}, t^n]}(\bm{\xi}) = \bm{\xi} - \Delta t I_{\Delta x} \left[  \sum_i a^l_i \tilde{\u} ( \tilde{\map}_{[t^{n+1}, t_i]}(\bm{\xi}), t_i  )   \right] \\
        & F^{n+1} = I_{\Delta x} \left[ \exp ( -\frac{\Delta t}{\tau} ) F^n \circ \map_{[t^{n+1}, t^n]} + \left(1 - \exp ( -\frac{\Delta t}{\tau} )  \right) \sum_i a^e_i M [ \tilde{f}(t_i) ] \circ \tilde{\map}_{[t^{n+1}, t_i]} \right] ,
    \end{aligned}
\end{equation}
where $I_{\Delta x}[\cdot]$ denotes Hermite cubic interpolation in space. Below is the implementation of an RK2 time-stepping scheme Algorithm \ref{algo:RK2} which we use as building block for our third order methods:
\begin{algorithm}[h]
  \small
  \caption{RK2 step}
  \label{algo:RK2}
  \begin{algorithmic}[1]
    \renewcommand{\algorithmicrequire}{\textbf{Input:}}
    \renewcommand{\algorithmicensure}{\textbf{Output:}}

    \REQUIRE Distribution $f^n$, time step $\Delta t$, Knudsen $\tau$
    \ENSURE $\map_{[t^{n+1}, t^n]}, f^{n+1}$

    \STATE $\u_1 \gets \u [f^n]$
    \STATE $\map_1  \gets \textit{id} - \Delta t \u^{(1)} \circ ( \textit{id} - \Delta t \u_1 ) $

    \STATE $f_2 \gets  \left( \exp ( -\frac{\Delta t}{\tau} ) f^n +  \left(1 - \exp ( -\frac{\Delta t}{\tau} )  \right) M[f^n] \right) \circ \map_1$

    \STATE $\u_2 \gets \u [f_2]$
    
    \STATE $\map_{[t^{n+1}, t^n]} \gets \frac{1}{2}(\map_1 + (\textit{id} - \Delta t \u_2 ) )$
    
    \STATE $f^{n+1} \gets  \exp ( -\frac{\Delta t}{\tau} ) f^n \circ \map_{[t^{n+1}, t^n]} + \frac{1}{2}\left(1 - \exp ( -\frac{\Delta t}{\tau} )  \right) \left( M[f^n] \circ \map_{[t^{n+1}, t^n]} + M[f_2] \right)$
  \end{algorithmic}
\end{algorithm}
Using the above RK2 scheme as the local time-extension operator, we can again use the quadrature \eqref{eqns:timestepScheme} to generate a RK3 scheme using quadrature and weights:
\begin{equation} \label{eqns:quadraturePts}
    \begin{aligned}
        & t_0 = t^n , & a^l_0 = \frac{1}{6}, \quad & a^e_0 = \frac{1}{6}, \\
        & t_1 = t^n + \frac{\Delta t}{2} , & a^l_1 = \frac{2}{3}, \quad & a^e_1 = 0, \\
        & t_2 = t^n + \frac{\Delta t}{2} (1 +  \exp( (t-t^n)/ \tau)) ,  & a^l_2 = 0, \quad &   a^e_2 = \frac{2}{3},\\
        & t_3 = t^{n+1} = t^n + \Delta t ,& a^l_3 = \frac{1}{6} , \quad & a^e_3 = \frac{1}{6}.
    \end{aligned}
\end{equation}
The pseudocode is given in Algorithm \ref{algo:expSL}.
\begin{algorithm}[h]
  \small
  \caption{SLstep -- semi-Lagrangian scheme with exponential coordinates}
  \label{algo:expSL}
  \begin{algorithmic}[1]
    \renewcommand{\algorithmicrequire}{\textbf{Input:}}
    \renewcommand{\algorithmicensure}{\textbf{Output:}}

    \REQUIRE Map and source term accumulation $\map^n, F^n$, distribution $f^n$, time step $\Delta t$, Knudsen $\tau$
    \ENSURE $\map^{n+1}, F^{n+1}$

    \STATE Let $t_i, a^l_i, a^e_i$ for $i=0, \ldots, 3$ as given in \eqref{eqns:quadraturePts}

    \STATE Let $f_0 = f^n$
    \FOR{$j=1, \ldots, 3$}
        \STATE $(f_j, \tilde{\map}_{[t_{j}, t_{j-1}]})  \gets \textbf{RK2}(f_{j-1}, t_j - t_{j-1})$
    \ENDFOR

    \STATE Define $\tilde{\u}_j = \u [ f_j ]$, $\tilde{M}_j = M[f_j]$. Let $\tilde{\map}_3 = \textit{id}$ the identity map.
    \FOR{$j=2, \ldots 0$}
        \STATE $\tilde{\map}_j \gets \tilde{\map}_{[t_{j+1}, t_j]} \circ \tilde{\map}_{j+1} $
    \ENDFOR

    \STATE $\map_{[t^{n+1}, t^n]} \gets \textit{id} - \Delta t I_{\Delta x} \left[  \sum_{j =0 }^3 a^l_j \tilde{\u}_j \circ \tilde{\map}_{j}  \right]$

    \STATE $F^{n+1} \gets I_{\Delta x} \left[  \exp ( -\Delta t / \tau ) F^n \circ \map_{[t^{n+1}, t^n]} + \left(1 - \exp ( -\Delta t / \tau )  \right) \sum_{j =0 }^3 a^e_j \tilde{M}_j \circ \tilde{\map}_{j}  \right]$
    \STATE $\map^{n+1} \gets I_{\Delta x} \left[ \map^n \circ \map_{[t^{n+1}, t^n]} \right]$
  \end{algorithmic}
\end{algorithm}
The intermediate approximations $\tilde{f}$ and $\tilde{\map}$ are computed using an RK2 scheme between the consecutive time steps $t_i$ which formally yields a local truncation error of $\Delta t^3$ on all terms in the quadrature \eqref{eqns:timestepScheme}. We then expect an error of $\mathcal{O} ( \Delta t^4 + \Delta x^4 )$ for $\map_{[t^{n+1}, t^n]}$, and an error of $\mathcal{O} ( \left(1 - \exp ( -\Delta t / \tau )  \right)  \Delta t^3 )$ for $F^{n+1} -  \exp ( -\Delta t / \tau ) F^n \circ \map_{[t^{n+1}, t^n]} $, which is asymptotically $\Delta t^4$. With a fourth order local truncation error, we expect to achieve third order convergence for this scheme.


\subsection{Source term integration using DIRK}
\label{ssec:CMM_ST_DIRK}

Once we decouple the advection from the source term integration as described in the previous section, exponential decay from the source term can also be resolved using implicit methods such as diagonally implicit Runge--Kutta (DIRK) schemes. Employing a technique used in \cite{santagati2011new}, we can significantly simplify the scheme in the case of the BGK term so that the implicit evaluations can in fact be computed explicitly without any iterative schemes.

We use Alexander's third order DIRK scheme whose Butcher tableau is lower triangular:
\begin{equation} \label{tbl:DIRK3}
\begin{array}{c|c}\vect{c} &A\\\hline &\vect {b}^{T} \\\end{array} =
\begin{array}{c| ccc}
1/2 &  \gamma  &  0 &  0 \\
(1+\gamma)/2 & (1- \gamma )/2 & \gamma & 0 \\
1 &  1-\delta-\gamma  & \delta & \gamma \\
\hline
 & 1-\delta-\gamma & \delta & \gamma \\
\end{array}
\qquad \text{with} \quad 
\begin{array}{c}
\gamma = 0.435866521508459,\\
\delta = 0.193914628359152
\end{array}
\end{equation} 

The $s[f]$ term is treated implicitly for all diagonal entries, the key realization which simplifies the method is that in this source term, $M[f]$ can be treated explicitly. Consider the stepping scheme between $t^n$ and $t^{n+1} = t^n + \Delta t$ given by
\begin{subequations}
\begin{align}
    & f_k = f^n\circ \map_{[t^n + c_k \Delta t, t^n ]} + \Delta t \sum_{j=1}^k a_{kj} s_j \circ \map_{[t^n + c_k \Delta t, t^n + c_j \Delta t]} , \\
    & s_j = s[ f_j ] = \tau^{-1} (  M[f_j] - f_j ), \\
    & f^{n+1} = f^n\circ \map_{[t^{n+1}, t^n ]} + \Delta t \sum_{j=1}^3 b_{j} s_j \circ \map_{[t^{n+1}, t^n + c_j \Delta t]} .
\end{align}
\end{subequations}

Computations for the $k^{th}$ stage value $f_k$ involves solving
\begin{equation} \label{eq:stage_k_value}
    f_k = f^n\circ \map_{[t^n + c_k \Delta t, t^n ]} + \Delta t \sum_{j=1}^k a_{kj} s_j = \tilde{f}_{k} + \Delta t a_{kk} s_k \circ \map_{[t^n + c_k \Delta t, t^n + c_k \Delta t]},
\end{equation}
where we collect all previous stages terms in $\tilde{f}_{k}:$ 
\begin{equation} 
    \tilde{f}_{k} = f^n\circ \map_{[t^n + c_k \Delta t, t^n ]} + \Delta t \sum_{j=1}^{k-1} a_{kj} s_j \circ \map_{[t^n + c_k \Delta t, t^n + c_j \Delta t]} .
\end{equation} 

Noting that $s_k = \tau^{-1} (  M[f_k] - f_k )$ is composed with $\map_{[t^n + c_k \Delta t, t^n + c_k \Delta t]} = \textit{id}$, the implicit equation for \eqref{eq:stage_k_value} is then
\begin{equation}
    \left( 1 + a_{kk} \frac{\Delta t}{\tau} \right) f_k - a_{kk} \frac{\Delta t}{\tau} M[f_k] = \tilde{f}_{k} .
\end{equation}
Here we make use of a property of the Maxwellian previously used in \cite{santagati2011new}, i.e., that for a given distribution $f$, for $\alpha + \beta = 1$, we have that $\alpha f + \beta M[f]$ has the same moments as $f$, and therefore $M[\alpha f + \beta M[f]] = M[f]$. Applying this to the above, we get that 
\begin{equation}
    M [f_k] = M[\tilde{f}_{k}],
\end{equation}
which we now substitute into \eqref{eq:stage_k_value}. This gives an implementation of the DIRK time stepping schemes which requires only explicit evaluations:
\begin{subequations}
\begin{align}
    & f_k = \left( 1 + a_{kk} \frac{\Delta t}{\tau} \right)^{-1} \left( \tilde{f}_{k} + a_{kk} \frac{\Delta t}{\tau} M[\tilde{f}_{k}] \right) \label{eqn:def_F} \\
    & \tilde{f}_{k} = f^n\circ \map_{[t^n + c_k \Delta t, t^n ]} + \Delta t \sum_{j=1}^{k-1} a_{kj} s_j \circ \map_{[t^n + c_k \Delta t, t^n + c_j \Delta t]} ,\label{eqn:def_tildeF} \\
    & s_j = \tau^{-1} (  M[f_j] - f_j ) . \label{eqn:def_S},
\end{align}
\end{subequations}
with the final integral approximation given by
\begin{equation} \label{eqn:def_int}
    f^{n+1} - f^n \circ \map_{[t^{n+1}, t^n]} \approx \Delta t \sum_{k=1}^m b_k s_k \circ \map_{[t^{n+1}, t^n + c_k \Delta t]} .
\end{equation}

The full time-stepping algorithm is summarized in the following pseudocode Algorithm \ref{algo:DIRK}. We again use a first RK2 pass to generate approximate solutions at the DIRK3 time stages followed by a DIRK3 solve for the actual time stepping.
\begin{algorithm}[h]
  \small
  \caption{SLstep -- semi-Lagrangian scheme using DIRK3}
  \label{algo:DIRK}
  \begin{algorithmic}[1]
    \renewcommand{\algorithmicrequire}{\textbf{Input:}}
    \renewcommand{\algorithmicensure}{\textbf{Output:}}

    \REQUIRE Map and source term accumulation $\map^n, F^n$, distribution $f^n$, time step $\Delta t$, Knudsen $\tau$
    \ENSURE $\map^{n+1}, F^{n+1}$

    \STATE Let $t_i = t^n + c_i \Delta t, a_{ij}, b_i$ for $i,j=1, \ldots, 3$ as given in \eqref{tbl:DIRK3}, $t_0 = t^n$, $f_0 = f^n$
    \FOR{$j=1, \ldots, 3$}
        \STATE $(f_j, \tilde{\map}_{[t_{j}, t_{j-1}]})  \gets \textbf{RK2}(f_{j-1}, t_j - t_{j-1})$
    \ENDFOR

    \STATE Define $\tilde{\u}_j = \u [ f_j ]$, $\tilde{M}_j = M[f_j]$. Let $\tilde{\map}_3 = \textit{id}$ the identity map.
    \FOR{$j=2, \ldots 0$}
        \STATE $\tilde{\map}_j \gets \tilde{\map}_{[t_{j+1}, t_j]} \circ \tilde{\map}_{j+1} $
    \ENDFOR

    \STATE Let $\map_{[t^{n+1}, t^n]} \gets \textit{id} - \Delta t I_{\Delta x} \left[  \sum_{i = 1 }^3 b_i \tilde{\u}_i \circ \tilde{\map}_{i}  \right]$

    \FOR{$k=1,\dots,3$}
        \STATE $\tilde{f}_k \gets f_0 \circ \map_{[t^{n+1}, t^n]} + \Delta t \sum_{j=1}^{k-1} a_{kj} \tau^{-1}( \tilde{M}_j - f_j ) \circ \tilde{\map}_j$
        \STATE $f_k \gets \left( 1 + a_{kk} \frac{\Delta t}{\tau} \right)^{-1} \left( \tilde{f}_{k} + a_{kk} \frac{\Delta t}{\tau} \tilde{M}_k \circ \tilde{\map}_k \right)$
    \ENDFOR

    \STATE $F^{n+1} \gets I_{\Delta x} \left[ \left( 1 + a_{kk} \frac{\Delta t}{\tau} \right)^{-1} \left( (F^n + b_3 \frac{\Delta t}{\tau} \tilde{M}_3 ) \circ \map_{[t^{n+1}, t^n]}  + \Delta t \sum_{i=1}^{k-1} b_i \tau^{-1}( \tilde{M}_i - f_i ) \circ \tilde{\map}_i  \right) \right]$
    \STATE $\map^{n+1} \gets I_{\Delta x} \left[ \map^n \circ \map_{[t^{n+1}, t^n]} \right]$
    
  \end{algorithmic}
\end{algorithm}


\subsection{CMM for Vlasov--Poisson--BGK}
\label{ssec:CMM_VPBGK}

In the full algorithm with CMM submap and source term decomposition, for a list of remapping times $0 = T_0 < T_1 < \cdots < T_m \leq t^n$, we have the following scheme. Given a list of submaps and accumulated source term pairs for the history: $\{ ( \map_{[T_i, T_{i-1}]} , F_{[T_i, T_{i-1}]} ) \}_{i=1}^m$ and the current submap and source $( \map_{[t^n, T_m]} , F_{[t^n, T_m]} )$, we evolve the current submap and source accumulation using a semi--Lagrangian step with either exponential coordinates (Algorithm \ref{algo:expSL}) or DIRK3 (Algorithm \ref{algo:DIRK}), and the distribution $f$ is given by \eqref{eqn:decomposedDistEval}. The full time-stepping algorithm with remapping is described in pseudocode Algorithm \ref{algo:CMM_VPBGK}.

\begin{algorithm}[h]
  \small
  \caption{CMM time stepping for VP--BGK}
  \label{algo:CMM_VPBGK}
  \begin{algorithmic}[1]
    \renewcommand{\algorithmicrequire}{\textbf{Input:}}
    \renewcommand{\algorithmicensure}{\textbf{Output:}}

    \REQUIRE $m = N_{\textit{sub}}$, $\Delta t$, $\tau$, time subdivision $0 = T_0 < T_1 < \cdots < T_m \leq t^n$, \\ history $\{ ( \map_{[T_i, T_{i-1}]} , F_{[T_i, T_{i-1}]} ) \}_{i=1}^m$, current state $( \map_{[t^n, T_m]} , F_{[t^n, T_m]} )$
    \ENSURE $\{ ( \map_{[T_i, T_{i-1}]} , F_{[T_i, T_{i-1}]} ) \}_{i=1}^m$, $( \map_{[t^{n+1}, T_m]} , F_{[t^{n+1}, T_m]} )$
    \STATE Define $f^n$ from \eqref{eqn:decomposedDistEval}
    \COMMENT{Terms with pre-factor $\exp ( -(t - T_i)/\tau )$ sufficiently small can be neglected from the evaluation.}
    \STATE $( \map_{[t^{n+1}, T_m]} , F_{[t^{n+1}, T_m]} ) \gets \text{SLstep}( \map_{[t^{n}, T_m]} , F_{[t^{n}, T_m]}, f^n ) $
    \IF{Trigger remapping}
        \STATE $T_{m+1} \gets t^{n+1}$
        \STATE $\map_{[T_{m+1}, T_m]} \gets \map_{[t^{n+1}, T_m]}$, $F_{[T_{m+1}, T_m]} \gets F_{[t^{n+1}, T_m]}$
        \STATE $\map_{[t^{n+1}, T_{m+1}]} \gets \textit{id}$, $F_{[t^{n+1}, T_{m+1}]} \gets 0$
        \STATE $N_{\textit{sub}} \gets N_{\textit{sub}} +1$
    \ENDIF
  \end{algorithmic}
\end{algorithm}

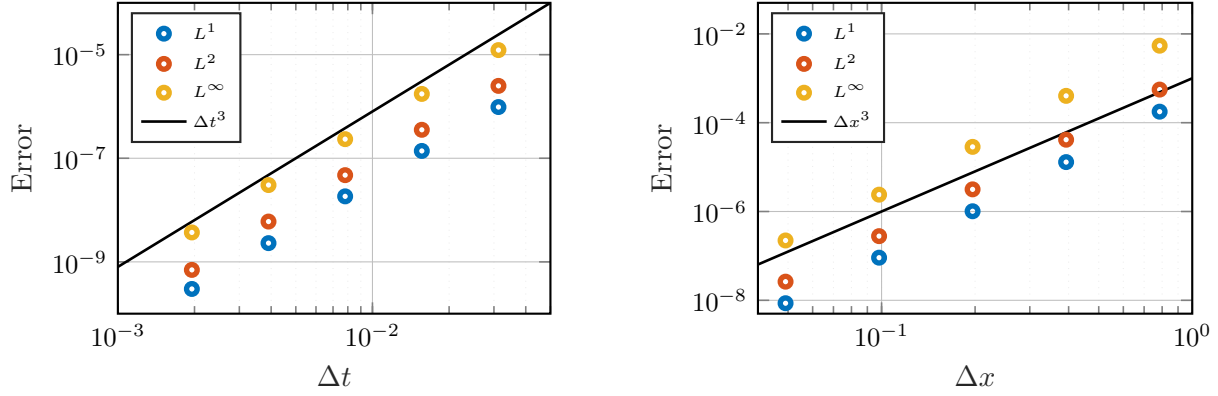
\begin{figure}[h]
    \centering
    \setlength\figureheight{0.25\textwidth}
    \setlength\figurewidth{0.35\textwidth}
%
\definecolor{mycolor1}{rgb}{0.00000,0.44700,0.74100}%
\definecolor{mycolor2}{rgb}{0.85000,0.32500,0.09800}%
\begin{tikzpicture}

\begin{axis}[%
width=0.98\figurewidth,
height=\figureheight,
at={(0\figurewidth,0\figureheight)},
scale only axis,
xmode=log,
xmin=0.0010000000,
xmax=0.0500000000,
xminorticks=true,
xlabel style={font=\color{white!15!black}},
xlabel={$\Delta t$},
ymode=log,
ymin=0.0000000008,
ymax=0.0001000000,
ytick={1e-09, 1e-07, 1e-05},
yminorticks=true,
ylabel style={font=\color{white!15!black}},
ylabel={$L^\infty$ error},
axis background/.style={fill=white},
xmajorgrids,
xminorgrids,
ymajorgrids,
yminorgrids,
legend style={at={(0.97,0.03)}, anchor=south east, legend cell align=left, align=left, draw=white!15!black},
scaled ticks=false, xticklabel style={/pgf/number format/fixed},yticklabel style={/pgf/number format/fixed}
]
\addplot [color=mycolor1, line width=3.0pt, only marks, mark=o, mark options={solid, mycolor1}]
  table[row sep=crcr]{%
0.0019531250	0.0000000037\\
0.0039062500	0.0000000301\\
0.0078125000	0.0000002306\\
0.0156250000	0.0000017214\\
0.0312500000	0.0000120866\\
};
\addlegendentry{Exp}

\addplot [color=mycolor2, line width=3.0pt, only marks, mark=o, mark options={solid, mycolor2}]
  table[row sep=crcr]{%
0.0019531250	0.0000000026\\
0.0039062500	0.0000000212\\
0.0078125000	0.0000001646\\
0.0156250000	0.0000012512\\
0.0312500000	0.0000090793\\
};
\addlegendentry{DIRK}

\addplot [color=black]
  table[row sep=crcr]{%
0.0019531250	0.0000000075\\
0.0039062500	0.0000000596\\
0.0078125000	0.0000004768\\
0.0156250000	0.0000038147\\
0.0312500000	0.0000305176\\
};
\addlegendentry{$\Delta t^3$}

\end{axis}

\begin{axis}[%
width=1.287\figurewidth,
height=1.287\figureheight,
at={(-0.185\figurewidth,-0.191\figureheight)},
scale only axis,
xmin=0.0000000000,
xmax=1.0000000000,
ymin=0.0000000000,
ymax=1.0000000000,
axis line style={draw=none},
ticks=none,
axis x line*=bottom,
axis y line*=left,
scaled ticks=false, xticklabel style={/pgf/number format/fixed},yticklabel style={/pgf/number format/fixed}
]
\end{axis}
\end{tikzpicture}%
    \hfill
%
\definecolor{mycolor1}{rgb}{0.00000,0.44700,0.74100}%
\definecolor{mycolor2}{rgb}{0.85000,0.32500,0.09800}%
\begin{tikzpicture}

\begin{axis}[%
width=0.98\figurewidth,
height=\figureheight,
at={(0\figurewidth,0\figureheight)},
scale only axis,
xmode=log,
xmin=0.0300000000,
xmax=1.5000000000,
xminorticks=true,
xlabel style={font=\color{white!15!black}},
xlabel={$\Delta x$},
ymode=log,
ymin=0.0000001000,
ymax=0.0125000000,
ytick={1e-07, 1e-05, 1e-03},
yminorticks=true,
ylabel style={font=\color{white!15!black}},
ylabel={$L^\infty$ error},
axis background/.style={fill=white},
xmajorgrids,
xminorgrids,
ymajorgrids,
yminorgrids,
legend style={at={(0.97,0.03)}, anchor=south east, legend cell align=left, align=left, draw=white!15!black},
scaled ticks=false, xticklabel style={/pgf/number format/fixed},yticklabel style={/pgf/number format/fixed}
]
\addplot [color=mycolor1, line width=3.0pt, only marks, mark=o, mark options={solid, mycolor1}]
  table[row sep=crcr]{%
0.7853981634	0.0075161894\\
0.3926990817	0.0005184871\\
0.1963495408	0.0000349471\\
0.0981747704	0.0000027769\\
0.0490873852	0.0000002436\\
};
\addlegendentry{Exp}

\addplot [color=mycolor2, line width=3.0pt, only marks, mark=o, mark options={solid, mycolor2}]
  table[row sep=crcr]{%
0.7853981634	0.0050580158\\
0.3926990817	0.0003283963\\
0.1963495408	0.0000258525\\
0.0981747704	0.0000022393\\
0.0490873852	0.0000002141\\
};
\addlegendentry{DIRK}

\addplot [color=black]
  table[row sep=crcr]{%
0.7853981634	0.0048447307\\
0.3926990817	0.0006055913\\
0.1963495408	0.0000756989\\
0.0981747704	0.0000094624\\
0.0490873852	0.0000011828\\
};
\addlegendentry{$\Delta x^3$}

\end{axis}

\begin{axis}[%
width=1.287\figurewidth,
height=1.287\figureheight,
at={(-0.185\figurewidth,-0.191\figureheight)},
scale only axis,
xmin=0.0000000000,
xmax=1.0000000000,
ymin=0.0000000000,
ymax=1.0000000000,
axis line style={draw=none},
ticks=none,
axis x line*=bottom,
axis y line*=left,
scaled ticks=false, xticklabel style={/pgf/number format/fixed},yticklabel style={/pgf/number format/fixed}
]
\end{axis}
\end{tikzpicture}%
    \caption{Error plots of the distribution functions for $\textit{Kn} = 0.1$, $t_{\textit{end}} = 1$ using CMM without remapping. Both spatial and temporal error go to $3^{rd}$ in the asymptotics. Error estimated by comparison against a reference test with $N_t = N_x = 1024$.}
    \label{fig:errorPlots}
\end{figure}

We tested this scheme by running the nonlinear Landau damping test (see section \ref{ssec:VPBGK}) with Knudsen number $0.1$ to a final time of $t=1$ without remappings. Figure \ref{fig:errorPlots} shows the convergence plot for this test obtained using self-comparison with a fine grid, our results support our estimate of $3^{rd}$ order global convergence in space and time.

\section{Numerical experiments}
\label{sec:pb}

In this section, we study the applicability of the CMM to kinetic equations with source terms by applying the method to several test cases studied in the literature. We first test the CMM using DIRK time-stepping (see section \ref{ssec:CMM_ST_DIRK}) to Sod's shock tube test in the case of the Boltzmann--BGK equations. Then the method using integrating factors (see section \ref{ssec:CMM_ST_exp}) is applied to the Vlasov--Poisson--BGK equations for the non-linear Landau damping and bump--on--tail instability test cases. We will also compare our results against a pseudo-spectral simulation as validation, a summary of the numerical implementation of the pseudo-spectral code is given in appendix \ref{sec:app2}. In all computations, the velocity domain $\Omega_v$ is taken to be $S^1$ instead of $\mathbb{R}$ so that the computational domain $\Omega$ is compact and that we do not need to implement boundary conditions in the velocity direction. This is achieved using the same domain periodization trick proposed in \cite{KrahYinNaveBergmannSchneider2024}.

\subsection{Boltzmann--BGK: Sod shock tube}
\label{ssec:BGK}

In the first example, we examine a pure BGK collision without the influence of the electric field. The test case is known as Sod's shock tube \cite{Sod1978} using the Boltzmann--BGK equations. Discontinuous equilibrium values describe the initial condition:
\begin{equation}
    \begin{cases}
        (\rho_\text{L}, u_\text{L}, p_\text{L}) = (1, 0 ,1) & \quad  \text{for } x < 0.5\\
        (\rho_\text{R}, u_\text{R}, p_\text{R}) = (0.125, 0 ,0.1) & \quad \text{for } x\ge 0.5.
    \end{cases}
\end{equation}
in a domain $x\in [0,1]$, which corresponds to a jump in density and pressure at $x=0.5$, a diaphragm which is removed at time $t=0$.

\begin{figure}[h]
  \centering
\includegraphics[width=0.45\textwidth]{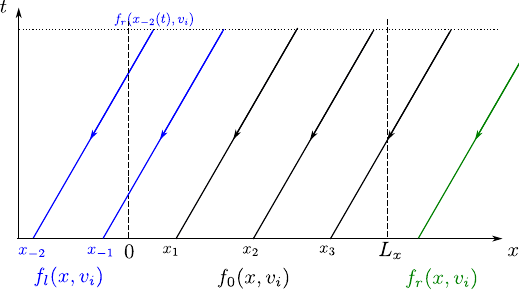}
    \caption{Dirichlet boundary conditions at $x=0$ and $x=L_x$.}
    \label{fig:BC_dirichlet}
\end{figure}

\begin{remark}[Dirichlet Boundary Conditions]
    In this test case, we may implement Dirichlet boundary conditions in the $x$-direction rather than the restrictive periodic boundary conditions in previous works on CMM. This allows us to impose in- and out-flow conditions. As shown in Fig.\ref{fig:BC_dirichlet}, for given velocity level $v_i$, the backward map $\map$ is constant in the $v$ component and a shift at rate $-v_i$ in $x$ direction, explicitly, $\map_{[t, 0]} (x, v) = (x - t v, v)$.
Thus for a given $v_i$ the characteristics are lines with constant slope in time. Following back the characteristic to its initial location, we assign it either the value $f_l$ if it exceeds the left boundary or $f_r$ in case it exceeds the right boundary.
\end{remark}

\begin{figure}[ht]
  \centering
  \setlength\figureheight{0.2\linewidth}
  \setlength\figurewidth{0.8\linewidth}
    \input{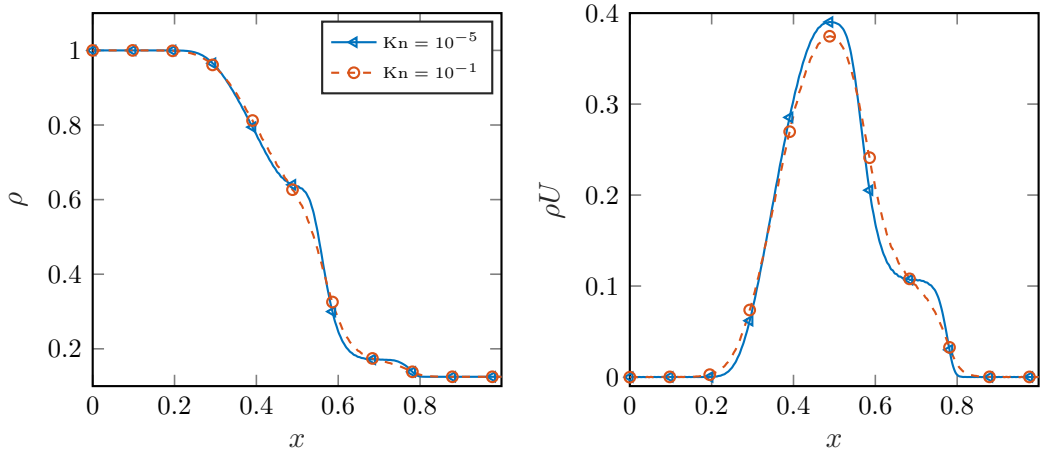}
    \caption{Sod shock tube: Macroscopic quantities $\rho,\rho u, \ekin$ at $t=0.12$ for $\Kn=0.01$ and $\Kn=10^{-5}$.}
    \label{fig:BGK-Macro}
\end{figure}

The time-stepping scheme in this test case uses the CMM approach with general source terms, integrating $s[f] = \frac{1}{\tau}(M[f] - f)$ directly using a third order DIRK scheme in time (see section \ref{ssec:CMM_ST_DIRK}). Here the DIRK approach is more general and can be applied to computations with boundary conditions, the method using exponential integration is more restrictive in the form of the equation. We study the system for two different Knudsen numbers $\Kn \in \{0.01,10^{-5}\}$ and plot the macroscopic quantities $\rho,\rho u$ and $\ekin$ in \cref{fig:BGK-Macro}. The results can be compared quantitatively to those given in \cite{KoellermeierKrahReissSchellin2024}.

\subsection{Vlasov--Poisson-BGK: Non-linear Landau damping}
\label{ssec:VPBGK}
In the second example, we simulate the non-linear Landau damping test case with BGK collision term (see \cite{LiuShiWanHeCao2020} as reference). This test case was already studied in \cite{KrahYinNaveBergmannSchneider2024} for the collisionless regime $\Kn\to\infty$, in this section, we will test the numerical scheme for the transitional regime for $\Kn \in \{ 0.1, 1, 10 \}$ and in the near-continuum regime for $\Kn = 10^{-4}$.

\begin{figure}[htp!]
  \centering
  \includegraphics[width=0.95\linewidth]{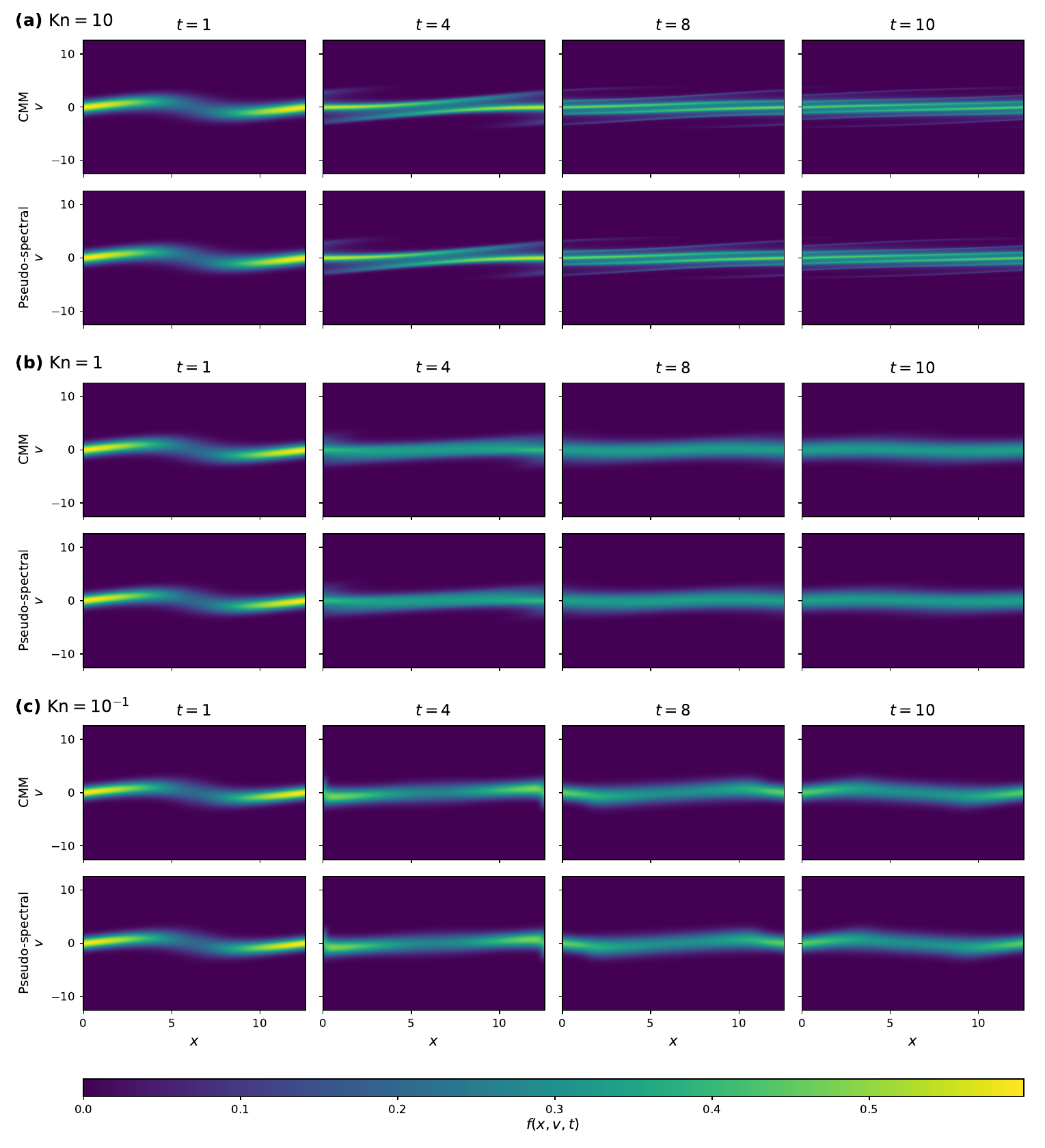}
  \caption{Non-linear Landau damping with $\alpha=0.5$, $k=0.5$, $(x,v)\in[0,L]\times[-4\pi,4\pi]$, $L=2\pi/k$, and $\tau=\Kn$: comparison of the particle distribution function for (a) $\Kn=10$, (b) $\Kn=1$, and (c) $\Kn=10^{-1}$. In each panel, the upper row shows the CMM result and the lower row shows the pseudo-spectral reference result at $t=1,4,8,$ and $10$. The CMM uses a $128^2$ map grid and a $512^2$ sampling grid; the pseudo-spectral reference uses a $512^2$ phase-space grid.}
  \label{fig:non-lin-BGK-PDF-compare}
\end{figure}

\begin{figure}[htp!]
  \centering
  \includegraphics[width=0.95\linewidth]{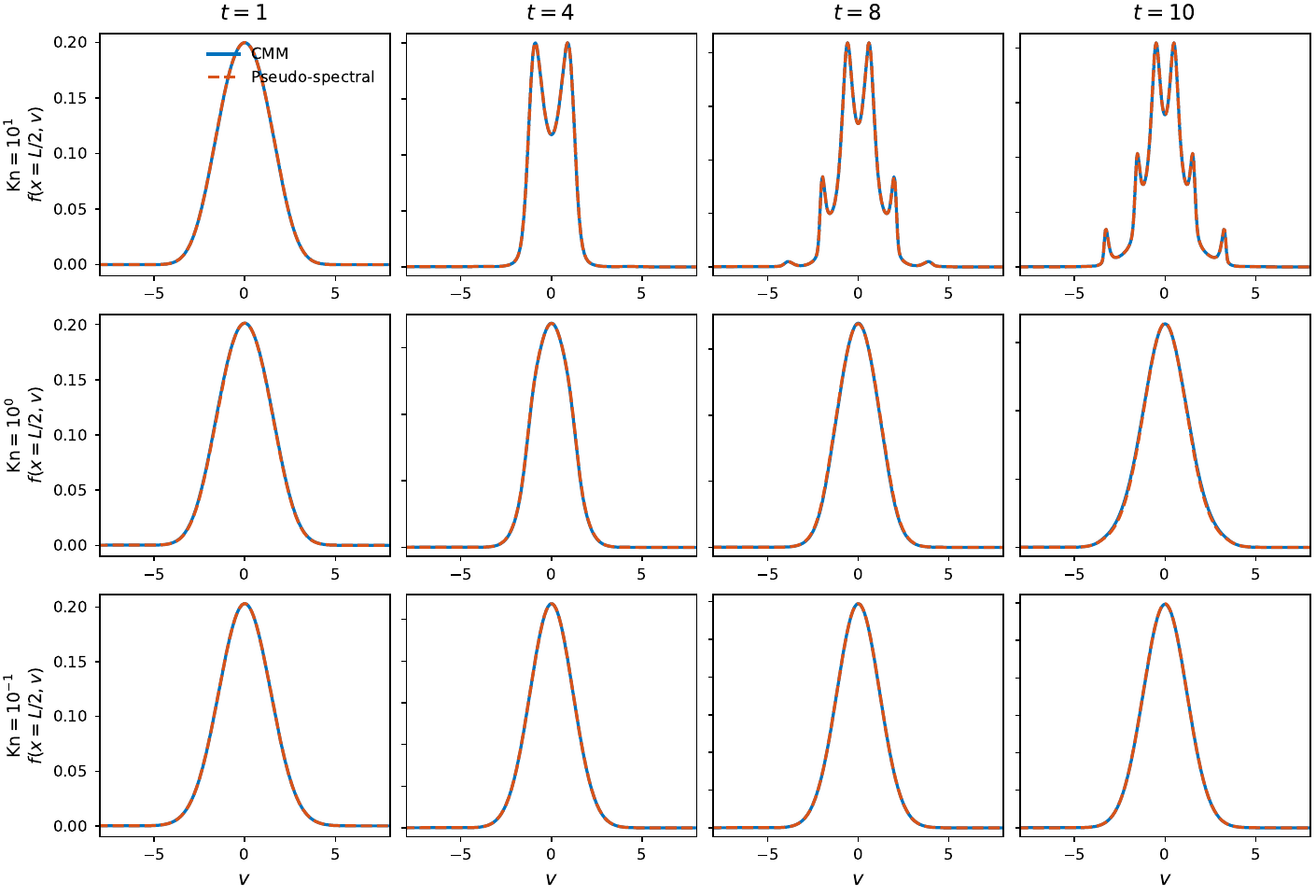}
  \caption{Non-linear Landau damping: corresponding velocity profiles $f(x=L/2,v)$ for $\Kn=10,1,$ and $10^{-1}$, using the same parameters and grids as Fig.~\ref{fig:non-lin-BGK-PDF-compare}. The CMM and pseudo-spectral methods are compared at $t=1,4,8,$ and $10$.}
  \label{fig:non-lin-BGK-profiles}
\end{figure}

\begin{figure}[htp!]
  \centering
  \includegraphics[width=0.95\linewidth]{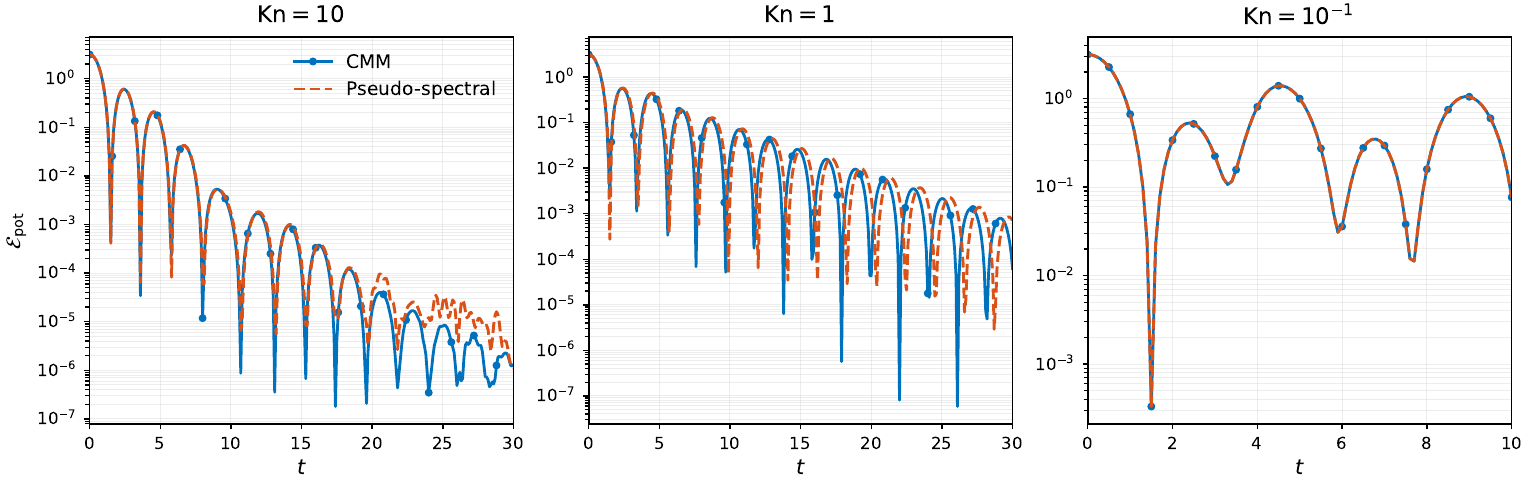}
  \caption{Non-linear Landau damping: electrostatic energy $\mathcal{E}_{\mathrm{pot}}$ for $\Kn=10,1,$ and $10^{-1}$, using the same parameters and grids as Fig.~\ref{fig:non-lin-BGK-PDF-compare}.}
  \label{fig:non-lin-BGK-epot}
\end{figure}

The initial distribution is defined by the following,
\begin{equation}
\label{eq:landau_damping_inicond}
    f_0(x,v) =  \left(1+\alpha\cos(kx) \right)\frac{1}{\sqrt{2\pi}} \, e^{-v^2/2}\,,
\end{equation}
with $\alpha = 0.5, k=0.5$ in the domain $[0,L]\times[-4\pi,4\pi]$, $L=2\pi/k$. We compare the results obtained from an implementation of CMM using the method described in section \ref{ssec:CMM_ST_exp} against a pseudo-spectral reference computation detailed in appendix \ref{sec:app2}. The CMM uses a $128^2$ map and source term grid and a $512^2$ sampling grid, while the pseudo-spectral reference uses a $512^2$ phase-space grid. Figure~\ref{fig:non-lin-BGK-PDF-compare} shows the phase-space distribution at selected times, Fig.~\ref{fig:non-lin-BGK-profiles} shows the corresponding velocity profiles at $x=L/2$, and Fig.~\ref{fig:non-lin-BGK-epot} compares the electrostatic energy. For $\Kn=10^{-1}$, the pseudo-spectral simulation could not be continued reliably beyond $t=10$ in our tests, as nonphysical negative values appeared in the distribution function and corrupted the moment evaluation in the BGK relaxation term. We therefore restrict the pseudo-spectral comparison for this case to $t\leq 10$.

\begin{figure}[ht]
{
\graphicspath{{figs/Kn1e-4}}

    \centering

    \begin{minipage}{0.63\textwidth}
        \centering
        \begin{subfigure}[t]{0.45\linewidth}
            \captionsetup{aboveskip=-8pt}
            \setlength\figureheight{0.6\linewidth}
            \setlength\figurewidth{0.9\linewidth}
%
\definecolor{mycolor1}{rgb}{0.12941,0.12941,0.12941}%
\begin{tikzpicture}

\begin{axis}[%
width=0.951\figurewidth,
height=\figureheight,
at={(0\figurewidth,0\figureheight)},
scale only axis,
axis on top,
xmin=-0.00,
xmax=12.566370614359172,
xtick={0.0000000000,6.2831853072,12.5663606144},
xticklabels={},
ymin=-12.5663706144,
ymax=12.5663706144,
ytick={-9.4247779608,0.0000000000,9.4247779608},
yticklabels={{$\text{-3}\pi$},{0},{$\text{3}\pi$}},
ylabel style={font=\color{mycolor1}},
ylabel={$v$},
axis background/.style={fill=white}
]
\addplot [forget plot] graphics [xmin=-0.0061359232, xmax=12.5602346912, ymin=-12.5786424607, ymax=12.5540987681] {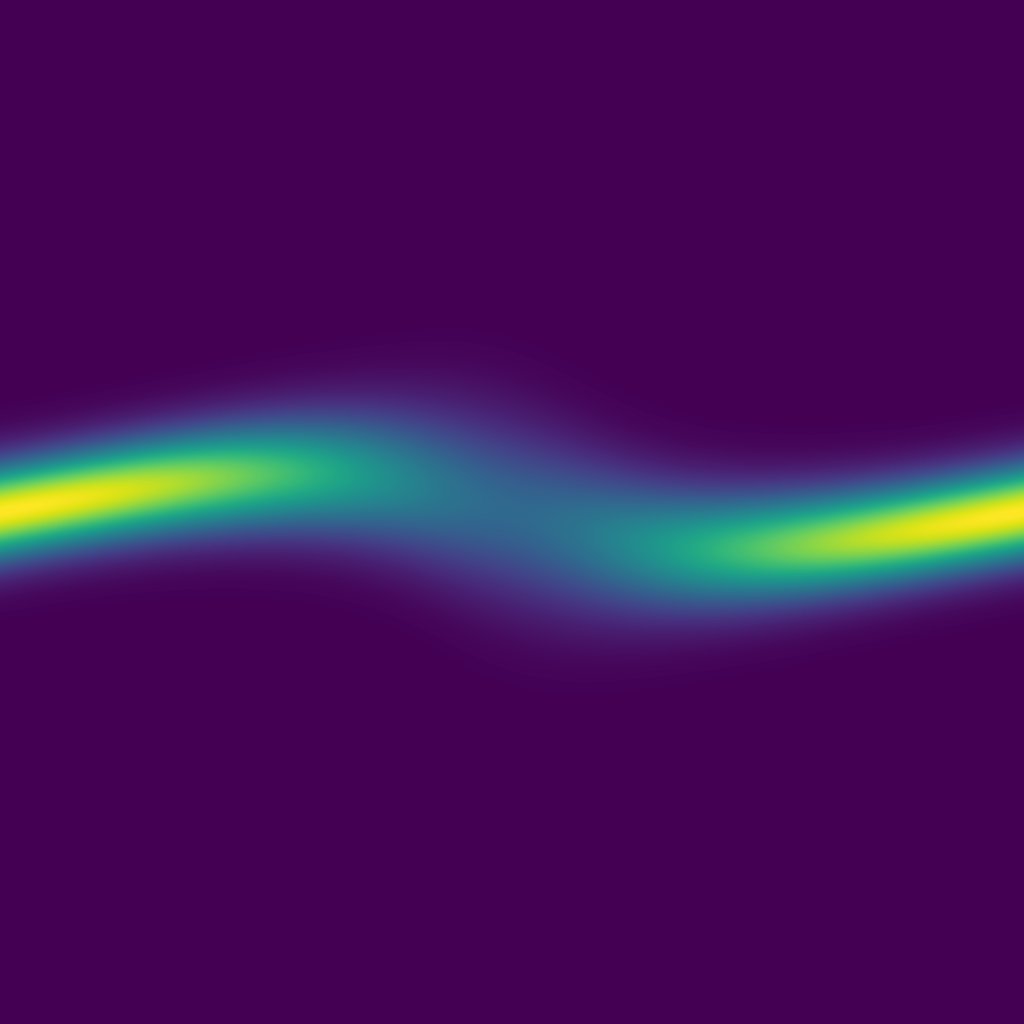};
\end{axis}
\end{tikzpicture}%
            \caption{$t=1$}
        \end{subfigure}
        \hfill
        \begin{subfigure}[t]{0.45\linewidth}
            \captionsetup{aboveskip=-8pt}
            \setlength\figureheight{0.6\linewidth}
            \setlength\figurewidth{0.9\linewidth}
%
\definecolor{mycolor1}{rgb}{0.12941,0.12941,0.12941}%
\begin{tikzpicture}

\begin{axis}[%
width=0.951\figurewidth,
height=\figureheight,
at={(0\figurewidth,0\figureheight)},
scale only axis,
axis on top,
xmin=-0.00,
xmax=12.566370614359172,
xtick={0.0000000000,6.2831853072,12.5663606144},
xticklabels={},
ymin=-12.5663706144,
ymax=12.5663706144,
ytick={-9.4247779608,0.0000000000,9.4247779608},
yticklabels={},
axis background/.style={fill=white}
]
\addplot [forget plot] graphics [xmin=-0.0061359232, xmax=12.5602346912, ymin=-12.5786424607, ymax=12.5540987681] {dist101-1.png};
\end{axis}
\end{tikzpicture}%
            \caption{$t=10$}
        \end{subfigure}

        \vspace{0.2em}

        \begin{subfigure}[t]{0.45\linewidth}
            \captionsetup{aboveskip=-2pt}
            \setlength\figureheight{0.6\linewidth}
            \setlength\figurewidth{0.9\linewidth}
%
\definecolor{mycolor1}{rgb}{0.12941,0.12941,0.12941}%
\begin{tikzpicture}

\begin{axis}[%
width=0.951\figurewidth,
height=\figureheight,
at={(0\figurewidth,0\figureheight)},
scale only axis,
axis on top,
xmin=-0.00,
xmax=12.566370614359172,
xtick={0.0000000000,6.2831853072,12.5663606144},
xticklabels={{0},{$\text{2}\pi$},{$\text{4}\pi$}},
xlabel style={font=\color{mycolor1}},
xlabel={$x$},
ymin=-12.5663706144,
ymax=12.5663706144,
ytick={-9.4247779608,0.0000000000,9.4247779608},
yticklabels={{$\text{-3}\pi$},{0},{$\text{3}\pi$}},
ylabel style={font=\color{mycolor1}},
ylabel={$v$},
axis background/.style={fill=white}
]
\addplot [forget plot] graphics [xmin=-0.0061359232, xmax=12.5602346912, ymin=-12.5786424607, ymax=12.5540987681] {dist201-1.png};
\end{axis}
\end{tikzpicture}%
            \caption{$t=20$}
        \end{subfigure}
        \hfill
        \begin{subfigure}[t]{0.45\linewidth}
            \captionsetup{aboveskip=-2pt}
            \setlength\figureheight{0.6\linewidth}
            \setlength\figurewidth{0.9\linewidth}
%
\definecolor{mycolor1}{rgb}{0.12941,0.12941,0.12941}%
\begin{tikzpicture}

\begin{axis}[%
width=0.951\figurewidth,
height=\figureheight,
at={(0\figurewidth,0\figureheight)},
scale only axis,
axis on top,
xmin=-0.00,
xmax=12.566370614359172,
xtick={0.0000000000,6.2831853072,12.5663606144},
xticklabels={{0},{$\text{2}\pi$},{$\text{4}\pi$}},
xlabel style={font=\color{mycolor1}},
xlabel={$x$},
ymin=-12.5663706144,
ymax=12.5663706144,
ytick={-9.4247779608,0.0000000000,9.4247779608},
yticklabels={},
axis background/.style={fill=white}
]
\addplot [forget plot] graphics [xmin=-0.0061359232, xmax=12.5602346912, ymin=-12.5786424607, ymax=12.5540987681] {dist301-1.png};
\end{axis}
\end{tikzpicture}%
            \caption{$t=30$}
        \end{subfigure}
    \end{minipage}
    \hfill
    \begin{minipage}{0.35\textwidth}
        \centering
        \begin{subfigure}[t]{\linewidth}
            \setlength\figureheight{0.6\linewidth}
            \setlength\figurewidth{0.8\linewidth}
%
\definecolor{mycolor1}{rgb}{0.06600,0.44300,0.74500}%
\definecolor{mycolor2}{rgb}{0.12941,0.12941,0.12941}%
\begin{tikzpicture}

\begin{axis}[%
width=0.951\figurewidth,
height=\figureheight,
at={(0\figurewidth,0\figureheight)},
scale only axis,
xmin=0.0000000000,
xmax=30.0000000000,
xlabel style={font=\color{mycolor2}},
xlabel={$t$},
ymode=log,
ymin=0.0009035626,
ymax=3.1415926534,
yminorticks=true,
ylabel style={font=\color{mycolor2}},
ylabel={$\mathcal{E}_{\textit{pot}}$},
axis background/.style={fill=white},
scaled ticks=false, xticklabel style={/pgf/number format/fixed},yticklabel style={/pgf/number format/fixed}
]
\addplot [color=mycolor1, forget plot]
  table[row sep=crcr]{%
0.0000000000	3.1415926534\\
0.1000000000	3.0962694335\\
0.2000000000	2.9753128636\\
0.3000000000	2.7853432708\\
0.4000000000	2.5366222544\\
0.5000000000	2.2423824324\\
0.6000000000	1.9179892966\\
0.7000000000	1.5799939449\\
0.8000000000	1.2451313171\\
0.9000000000	0.9293113099\\
1.0000000000	0.6466452828\\
1.1000000000	0.4085534376\\
1.2000000000	0.2230125178\\
1.3000000000	0.0940240409\\
1.4000000000	0.0213949492\\
1.5000000000	0.0009035626\\
1.6000000000	0.0248657185\\
1.7000000000	0.0830372556\\
1.8000000000	0.1637232424\\
1.9000000000	0.2549362125\\
2.0000000000	0.3454581802\\
2.1000000000	0.4257010848\\
2.2000000000	0.4883097006\\
2.3000000000	0.5284958116\\
2.4000000000	0.5441251519\\
2.5000000000	0.5355976737\\
2.6000000000	0.5055692723\\
2.7000000000	0.4585620198\\
2.8000000000	0.4005055385\\
2.9000000000	0.3382425314\\
3.0000000000	0.2790273461\\
3.1000000000	0.2300404177\\
3.2000000000	0.1979407875\\
3.3000000000	0.1884682352\\
3.4000000000	0.2061086707\\
3.5000000000	0.2538342598\\
3.6000000000	0.3329160100\\
3.7000000000	0.4428195776\\
3.8000000000	0.5811771876\\
3.9000000000	0.7438358675\\
4.0000000000	0.9249390853\\
4.1000000000	1.1125741553\\
4.2000000000	1.2845723799\\
4.3000000000	1.4233770688\\
4.4000000000	1.5187208985\\
4.5000000000	1.5659998849\\
4.6000000000	1.5649670328\\
4.7000000000	1.5186844193\\
4.8000000000	1.4326677412\\
4.9000000000	1.3141645554\\
5.0000000000	1.1715229658\\
5.1000000000	1.0136245722\\
5.2000000000	0.8493701901\\
5.3000000000	0.6872170724\\
5.4000000000	0.5347809863\\
5.5000000000	0.3984890562\\
5.6000000000	0.2832711511\\
5.7000000000	0.1924699865\\
5.8000000000	0.1277254499\\
5.9000000000	0.0890025048\\
6.0000000000	0.0747172235\\
6.1000000000	0.0819459791\\
6.2000000000	0.1066953709\\
6.3000000000	0.1442077547\\
6.4000000000	0.1892625043\\
6.5000000000	0.2356153199\\
6.6000000000	0.2744880259\\
6.7000000000	0.2983987454\\
6.8000000000	0.3038018830\\
6.9000000000	0.2906472257\\
7.0000000000	0.2615405332\\
7.1000000000	0.2209860352\\
7.2000000000	0.1747128205\\
7.3000000000	0.1290691749\\
7.4000000000	0.0904802872\\
7.5000000000	0.0649711963\\
7.6000000000	0.0577589036\\
7.7000000000	0.0729177137\\
7.8000000000	0.1131208787\\
7.9000000000	0.1794602308\\
8.0000000000	0.2713438398\\
8.1000000000	0.3864699483\\
8.2000000000	0.5208736306\\
8.3000000000	0.6690405323\\
8.4000000000	0.8240778933\\
8.5000000000	0.9779248857\\
8.6000000000	1.1216057854\\
8.7000000000	1.2456957946\\
8.8000000000	1.3410792592\\
8.9000000000	1.3991103049\\
9.0000000000	1.4137824932\\
9.1000000000	1.3841393847\\
9.2000000000	1.3135639282\\
9.3000000000	1.2083916248\\
9.4000000000	1.0767678584\\
9.5000000000	0.9277615530\\
9.6000000000	0.7706621778\\
9.7000000000	0.6144074037\\
9.8000000000	0.4671117403\\
9.9000000000	0.3356825520\\
10.0000000000	0.2255201234\\
10.1000000000	0.1403047150\\
10.2000000000	0.0818769402\\
10.3000000000	0.0502185912\\
10.4000000000	0.0435389815\\
10.5000000000	0.0584666089\\
10.6000000000	0.0903377456\\
10.7000000000	0.1335636976\\
10.8000000000	0.1820528408\\
10.9000000000	0.2296351285\\
11.0000000000	0.2704249227\\
11.1000000000	0.2991039308\\
11.2000000000	0.3114079254\\
11.3000000000	0.3052202823\\
11.4000000000	0.2812344786\\
11.5000000000	0.2426637649\\
11.6000000000	0.1945478684\\
11.7000000000	0.1430298931\\
11.8000000000	0.0947050352\\
11.9000000000	0.0560562259\\
12.0000000000	0.0329749132\\
12.1000000000	0.0303641268\\
12.2000000000	0.0518221651\\
12.3000000000	0.0994058674\\
12.4000000000	0.1734721623\\
12.5000000000	0.2725958251\\
12.6000000000	0.3935613526\\
12.7000000000	0.5314301469\\
12.8000000000	0.6796930783\\
12.9000000000	0.8305260761\\
13.0000000000	0.9751529537\\
13.1000000000	1.1043148964\\
13.2000000000	1.2089687705\\
13.3000000000	1.2813222618\\
13.4000000000	1.3158334671\\
13.5000000000	1.3099036195\\
13.6000000000	1.2642892472\\
13.7000000000	1.1828869044\\
13.8000000000	1.0719970889\\
13.9000000000	0.9394703570\\
14.0000000000	0.7939306771\\
14.1000000000	0.6441157570\\
14.2000000000	0.4983274674\\
14.3000000000	0.3639799541\\
14.4000000000	0.2472377333\\
14.5000000000	0.1527416261\\
14.6000000000	0.0834244287\\
14.7000000000	0.0404202754\\
14.8000000000	0.0230716330\\
14.9000000000	0.0290357000\\
15.0000000000	0.0544875253\\
15.1000000000	0.0944101397\\
15.2000000000	0.1429502037\\
15.3000000000	0.1937908251\\
15.4000000000	0.2404971498\\
15.5000000000	0.2770327583\\
15.6000000000	0.2986191876\\
15.7000000000	0.3025346155\\
15.8000000000	0.2884740186\\
15.9000000000	0.2584423600\\
16.0000000000	0.2163366110\\
16.1000000000	0.1673897848\\
16.2000000000	0.1175969382\\
16.3000000000	0.0731485859\\
16.4000000000	0.0399612016\\
16.5000000000	0.0232586205\\
16.6000000000	0.0272199123\\
16.7000000000	0.0547059619\\
16.8000000000	0.1070601448\\
16.9000000000	0.1839819339\\
17.0000000000	0.2834711636\\
17.1000000000	0.4018406648\\
17.2000000000	0.5337981282\\
17.3000000000	0.6726051430\\
17.4000000000	0.8103272747\\
17.5000000000	0.9382075168\\
17.6000000000	1.0472829796\\
17.7000000000	1.1293246189\\
17.8000000000	1.1778883325\\
17.9000000000	1.1893258375\\
18.0000000000	1.1633075220\\
18.1000000000	1.1025595241\\
18.2000000000	1.0121863657\\
18.3000000000	0.8989230904\\
18.4000000000	0.7704437761\\
18.5000000000	0.6347545775\\
18.6000000000	0.4996711473\\
18.7000000000	0.3723756090\\
18.8000000000	0.2590501813\\
18.9000000000	0.1645876081\\
19.0000000000	0.0923810633\\
19.1000000000	0.0441976265\\
19.2000000000	0.0201395780\\
19.3000000000	0.0186964550\\
19.4000000000	0.0368878218\\
19.5000000000	0.0704917038\\
19.6000000000	0.1143456876\\
19.7000000000	0.1626914098\\
19.8000000000	0.2094968606\\
19.9000000000	0.2487613171\\
20.0000000000	0.2751716914\\
20.1000000000	0.2851498284\\
20.2000000000	0.2775431494\\
20.3000000000	0.2536304989\\
20.4000000000	0.2166968455\\
20.5000000000	0.1714568124\\
20.6000000000	0.1234698512\\
20.7000000000	0.0786015870\\
20.8000000000	0.0425488018\\
20.9000000000	0.0204327826\\
21.0000000000	0.0164621116\\
21.1000000000	0.0336651474\\
21.2000000000	0.0736921491\\
21.3000000000	0.1366865381\\
21.4000000000	0.2212241981\\
21.5000000000	0.3243194867\\
21.6000000000	0.4415211467\\
21.7000000000	0.5670072920\\
21.8000000000	0.6938265713\\
21.9000000000	0.8142034194\\
22.0000000000	0.9199905575\\
22.1000000000	1.0033788489\\
22.2000000000	1.0577629812\\
22.3000000000	1.0786526500\\
22.4000000000	1.0645285988\\
22.5000000000	1.0169984173\\
22.6000000000	0.9402594087\\
22.7000000000	0.8403394187\\
22.8000000000	0.7243657913\\
22.9000000000	0.5999279258\\
23.0000000000	0.4745374526\\
23.1000000000	0.3551773747\\
23.2000000000	0.2479327076\\
23.3000000000	0.1576990903\\
23.4000000000	0.0879693193\\
23.5000000000	0.0406999992\\
23.6000000000	0.0162613600\\
23.7000000000	0.0134727593\\
23.8000000000	0.0297243368\\
23.9000000000	0.0611814823\\
24.0000000000	0.1030626719\\
24.1000000000	0.1499711391\\
24.2000000000	0.1962445718\\
24.3000000000	0.2362980846\\
24.4000000000	0.2650766065\\
24.5000000000	0.2787827225\\
24.6000000000	0.2756210250\\
24.7000000000	0.2560946604\\
24.8000000000	0.2227904140\\
24.9000000000	0.1798828136\\
25.0000000000	0.1325603295\\
25.1000000000	0.0864732804\\
25.2000000000	0.0472400599\\
25.3000000000	0.0200223606\\
25.4000000000	0.0091714847\\
25.5000000000	0.0179455733\\
25.6000000000	0.0482971809\\
25.7000000000	0.1007305346\\
25.8000000000	0.1742277664\\
25.9000000000	0.2662437942\\
26.0000000000	0.3727710332\\
26.1000000000	0.4884780488\\
26.2000000000	0.6069292595\\
26.3000000000	0.7208968905\\
26.4000000000	0.8227978964\\
26.5000000000	0.9053170266\\
26.6000000000	0.9621795632\\
26.7000000000	0.9888989097\\
26.8000000000	0.9834266064\\
26.9000000000	0.9464747154\\
27.0000000000	0.8813398698\\
27.1000000000	0.7930820261\\
27.2000000000	0.6881175674\\
27.3000000000	0.5735037374\\
27.4000000000	0.4563903106\\
27.5000000000	0.3435545802\\
27.6000000000	0.2410165684\\
27.7000000000	0.1537316512\\
27.8000000000	0.0853599542\\
27.9000000000	0.0381137861\\
28.0000000000	0.0126853778\\
28.1000000000	0.0082570793\\
28.2000000000	0.0225947876\\
28.3000000000	0.0522224349\\
28.4000000000	0.0926702314\\
28.5000000000	0.1387806605\\
28.6000000000	0.1850462305\\
28.7000000000	0.2259788105\\
28.8000000000	0.2566043377\\
28.9000000000	0.2731265339\\
29.0000000000	0.2735414851\\
29.1000000000	0.2579229753\\
29.2000000000	0.2283122740\\
29.3000000000	0.1883364576\\
29.4000000000	0.1427095407\\
29.5000000000	0.0967187261\\
29.6000000000	0.0557463687\\
29.7000000000	0.0248485591\\
29.8000000000	0.0083979128\\
29.9000000000	0.0097930191\\
30.0000000000	0.0312352074\\
};
\end{axis}
\end{tikzpicture}%
            \caption{Potential energy}
        \end{subfigure}
    \end{minipage}

    \caption{Distribution functions and potential energy evolution for Knudsen number $\tau = 10^{-4}$ with map grid $256^2$, sampling grid $512^2$ and CFL = 2. In this case, $\Delta t \approx 0.1 \gg \tau$ indicating that the stiffness of the source term does not cause numerical instabilities.}
    \label{fig:Kn1e-4}
    }
\end{figure}
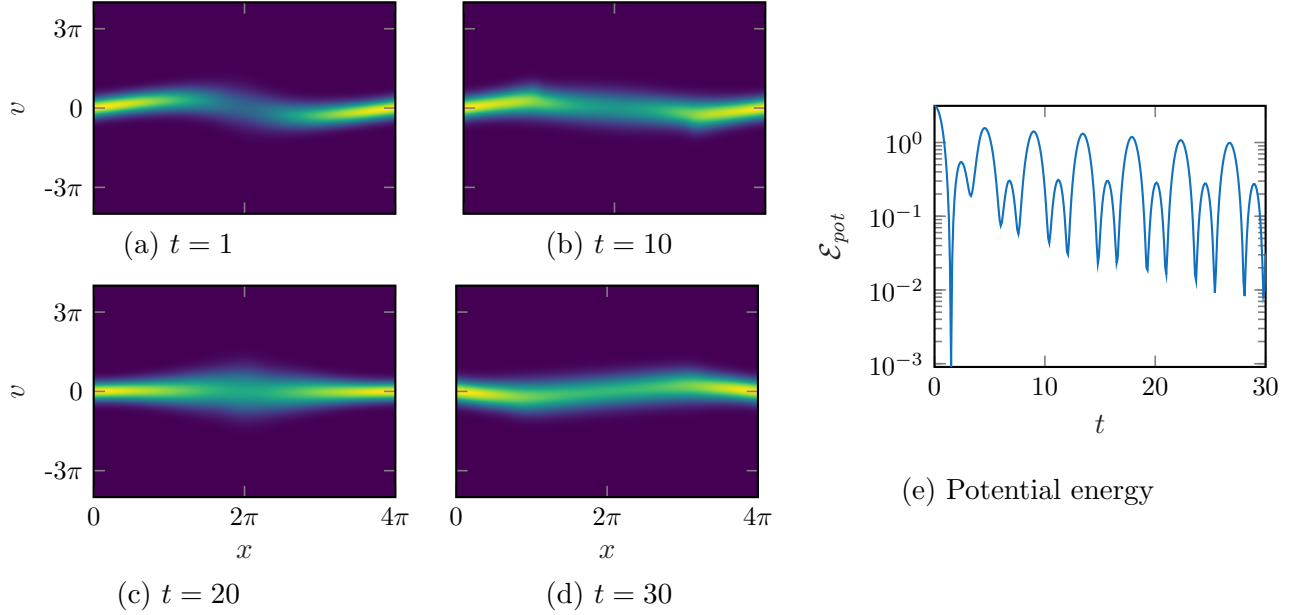

We also simulated the $\Kn = 10^{-4}$ case using the CMM to test the ability of the time-stepping scheme to handle the stiffness from the BGK source term. For this test, we fixed a map grid of $256^2$ with a sampling grid of $512^2$, the time step $\Delta t$ is chosen from a CFL condition independent of the Knudsen number. The results shown in Fig.~\ref{fig:Kn1e-4} indicate that the method shows no sign of instability despite the time step being much larger than the relaxation parameter.

\subsection{Vlasov--Poisson--BGK: Bump--on--Tail instability}
The last example we study is the bump--on--tail instability in the Vlasov--Poisson--BGK case. The initial condition, taken from \cite{LiuShiWanHeCao2020}, is given by
\begin{equation}
    f_0(x, v) = (1 + \alpha \cos (k x) ) \left( \frac{9}{10\sqrt{2 \pi}} e^{-0.5 v^2} + \frac{2}{10\sqrt{2 \pi}} e^{- 2(v-4.5)^2} \right) ,
\end{equation}
on a periodic spatial domain $\Omega_x = [0, 2 \pi/k]$.

The initial state (see Fig.~\ref{fig:bumptail_init}) is similar to the nonlinear Landau damping case with an added perturbation on the distribution profile in the velocity direction: a bump on the distribution tail. For sufficiently large Knudsen numbers, this perturbation produces oscillations in the large scale flow.

\begin{figure}[h]
    \centering
    {
    \graphicspath{{figs/BumpOnTail_Kn1e3}}
    \begin{subfigure}[t]{0.19\linewidth}
        \captionsetup{aboveskip=-2pt}
        \setlength\figureheight{0.7\linewidth}
        \setlength\figurewidth{1.\linewidth}
%
\begin{tikzpicture}

\begin{axis}[%
width=0.951\figurewidth,
height=\figureheight,
at={(0\figurewidth,0\figureheight)},
scale only axis,
axis on top,
xmin=0.0000000000,
xmax=20.9439510239,
xtick={\empty},
ymin=-3.1415926536,
ymax=9.4247779608,
ytick={\empty},
axis background/.style={fill=white}
]
\addplot [forget plot] graphics [xmin=-0.0102265386, xmax=20.9337244853, ymin=-12.5786424607, ymax=12.5540987681] {dist101-1.png};
\end{axis}
\end{tikzpicture}%
        \caption{$t=10$}
    \end{subfigure}
    \hfill
    \begin{subfigure}[t]{0.19\linewidth}
        \captionsetup{aboveskip=-2pt}
        \setlength\figureheight{0.7\linewidth}
        \setlength\figurewidth{1.\linewidth}
%
\begin{tikzpicture}

\begin{axis}[%
width=0.951\figurewidth,
height=\figureheight,
at={(0\figurewidth,0\figureheight)},
scale only axis,
axis on top,
xmin=0.0000000000,
xmax=20.9439510239,
xtick={\empty},
ymin=-3.1415926536,
ymax=9.4247779608,
ytick={\empty},
axis background/.style={fill=white}
]
\addplot [forget plot] graphics [xmin=-0.0102265386, xmax=20.9337244853, ymin=-12.5786424607, ymax=12.5540987681] {dist201-1.png};
\end{axis}
\end{tikzpicture}%
        \caption{$t=20$}
    \end{subfigure}
    \hfill
    \begin{subfigure}[t]{0.19\linewidth}
        \captionsetup{aboveskip=-2pt}
        \setlength\figureheight{0.7\linewidth}
        \setlength\figurewidth{1.\linewidth}
%
\begin{tikzpicture}

\begin{axis}[%
width=0.951\figurewidth,
height=\figureheight,
at={(0\figurewidth,0\figureheight)},
scale only axis,
axis on top,
xmin=0.0000000000,
xmax=20.9439510239,
xtick={\empty},
ymin=-3.1415926536,
ymax=9.4247779608,
ytick={\empty},
axis background/.style={fill=white}
]
\addplot [forget plot] graphics [xmin=-0.0102265386, xmax=20.9337244853, ymin=-12.5786424607, ymax=12.5540987681] {dist301-1.png};
\end{axis}
\end{tikzpicture}%
        \caption{$t=30$}
    \end{subfigure}
    \hfill
    \begin{subfigure}[t]{0.19\linewidth}
        \captionsetup{aboveskip=-2pt}
        \setlength\figureheight{0.7\linewidth}
        \setlength\figurewidth{1.\linewidth}
%
\begin{tikzpicture}

\begin{axis}[%
width=0.951\figurewidth,
height=\figureheight,
at={(0\figurewidth,0\figureheight)},
scale only axis,
axis on top,
xmin=0.0000000000,
xmax=20.9439510239,
xtick={\empty},
ymin=-3.1415926536,
ymax=9.4247779608,
ytick={\empty},
axis background/.style={fill=white}
]
\addplot [forget plot] graphics [xmin=-0.0102265386, xmax=20.9337244853, ymin=-12.5786424607, ymax=12.5540987681] {dist401-1.png};
\end{axis}
\end{tikzpicture}%
        \caption{$t=40$}
    \end{subfigure}
    \hfill
    \begin{subfigure}[t]{0.19\linewidth}
        \captionsetup{aboveskip=-2pt}
        \setlength\figureheight{0.7\linewidth}
        \setlength\figurewidth{1.\linewidth}
%
\begin{tikzpicture}

\begin{axis}[%
width=0.951\figurewidth,
height=\figureheight,
at={(0\figurewidth,0\figureheight)},
scale only axis,
axis on top,
xmin=0.0000000000,
xmax=20.9439510239,
xtick={\empty},
ymin=-3.1415926536,
ymax=9.4247779608,
ytick={\empty},
axis background/.style={fill=white}
]
\addplot [forget plot] graphics [xmin=-0.0102265386, xmax=20.9337244853, ymin=-12.5786424607, ymax=12.5540987681] {dist501-1.png};
\end{axis}
\end{tikzpicture}%
        \caption{$t=50$}
    \end{subfigure}
    \hfill
    }
    \caption{Time evolution of the distribution function for the bump--on--tail test case with Knudsen number $\Kn = 10^3$. Map and source term grid sizes $256^2$, sampling grid $512^2$.}
    \label{fig:bumptail_Kn1e3}
\end{figure}
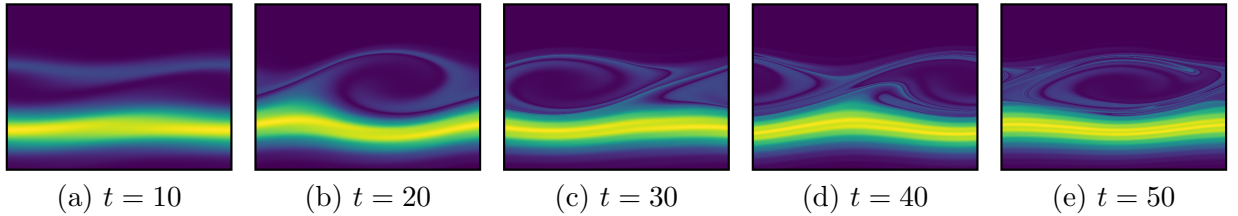

We show here the time evolution of the distribution functions for Knudsen numbers $\Kn = 10^2$ (Fig.~\ref{fig:bumptail_Kn1e2}) and $\Kn = 10^3$ (Fig.~\ref{fig:bumptail_Kn1e3}). The computations were performed using CMM with exponential integration of the source terms on a map and source term grid of $256^2$ and sampling grid of $512^2$ with $\Delta t \approx 0.1636$ corresponding to a CFL condition of $2$.

\begin{figure}[ht]
    \centering
    {
    \graphicspath{{figs/BumpOnTail_Kn1e2}}
    \setlength\figureheight{0.15\linewidth}
    \setlength\figurewidth{0.2\linewidth}
    \input{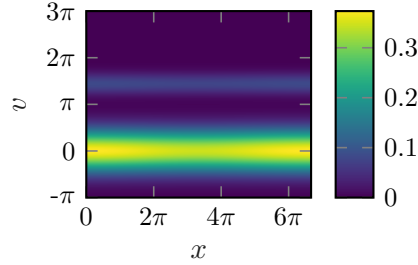}
    \caption{Initial distribution  for the bump-on-tail instability.}
    \label{fig:bumptail_init}
    }
\end{figure}
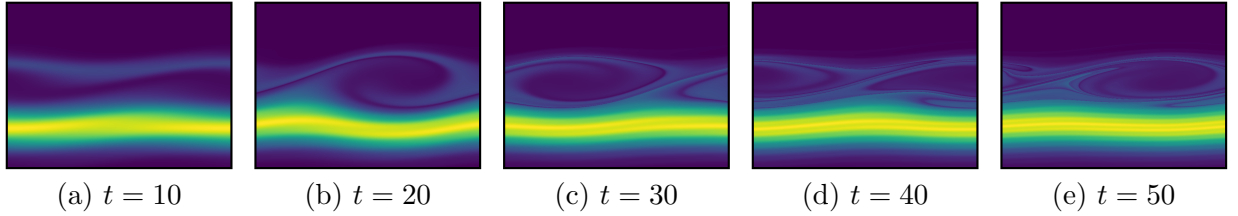
\begin{figure}[h]
    \centering
    {
    \graphicspath{{figs/BumpOnTail_Kn1e2}}
    \begin{subfigure}[t]{0.19\linewidth}
        \captionsetup{aboveskip=-2pt}
        \setlength\figureheight{0.7\linewidth}
        \setlength\figurewidth{1.\linewidth}
%
\begin{tikzpicture}

\begin{axis}[%
width=0.951\figurewidth,
height=\figureheight,
at={(0\figurewidth,0\figureheight)},
scale only axis,
axis on top,
xmin=0.0000000000,
xmax=20.9439510239,
xtick={\empty},
ymin=-3.1415926536,
ymax=9.4247779608,
ytick={\empty},
axis background/.style={fill=white}
]
\addplot [forget plot] graphics [xmin=-0.0102265386, xmax=20.9337244853, ymin=-12.5786424607, ymax=12.5540987681] {dist101-1.png};
\end{axis}
\end{tikzpicture}%
        \caption{$t=10$}
    \end{subfigure}
    \hfill
    \begin{subfigure}[t]{0.19\linewidth}
        \captionsetup{aboveskip=-2pt}
        \setlength\figureheight{0.7\linewidth}
        \setlength\figurewidth{1.\linewidth}
%
\begin{tikzpicture}

\begin{axis}[%
width=0.951\figurewidth,
height=\figureheight,
at={(0\figurewidth,0\figureheight)},
scale only axis,
axis on top,
xmin=0.0000000000,
xmax=20.9439510239,
xtick={\empty},
ymin=-3.1415926536,
ymax=9.4247779608,
ytick={\empty},
axis background/.style={fill=white}
]
\addplot [forget plot] graphics [xmin=-0.0102265386, xmax=20.9337244853, ymin=-12.5786424607, ymax=12.5540987681] {dist201-1.png};
\end{axis}
\end{tikzpicture}%
        \caption{$t=20$}
    \end{subfigure}
    \hfill
    \begin{subfigure}[t]{0.19\linewidth}
        \captionsetup{aboveskip=-2pt}
        \setlength\figureheight{0.7\linewidth}
        \setlength\figurewidth{1.\linewidth}
%
\begin{tikzpicture}

\begin{axis}[%
width=0.951\figurewidth,
height=\figureheight,
at={(0\figurewidth,0\figureheight)},
scale only axis,
axis on top,
xmin=0.0000000000,
xmax=20.9439510239,
xtick={\empty},
ymin=-3.1415926536,
ymax=9.4247779608,
ytick={\empty},
axis background/.style={fill=white}
]
\addplot [forget plot] graphics [xmin=-0.0102265386, xmax=20.9337244853, ymin=-12.5786424607, ymax=12.5540987681] {dist301-1.png};
\end{axis}
\end{tikzpicture}%
        \caption{$t=30$}
    \end{subfigure}
    \hfill
    \begin{subfigure}[t]{0.19\linewidth}
        \captionsetup{aboveskip=-2pt}
        \setlength\figureheight{0.7\linewidth}
        \setlength\figurewidth{1.\linewidth}
%
\begin{tikzpicture}

\begin{axis}[%
width=0.951\figurewidth,
height=\figureheight,
at={(0\figurewidth,0\figureheight)},
scale only axis,
axis on top,
xmin=0.0000000000,
xmax=20.9439510239,
xtick={\empty},
ymin=-3.1415926536,
ymax=9.4247779608,
ytick={\empty},
axis background/.style={fill=white}
]
\addplot [forget plot] graphics [xmin=-0.0102265386, xmax=20.9337244853, ymin=-12.5786424607, ymax=12.5540987681] {dist401-1.png};
\end{axis}
\end{tikzpicture}%
        \caption{$t=40$}
    \end{subfigure}
    \hfill
    \begin{subfigure}[t]{0.19\linewidth}
        \captionsetup{aboveskip=-2pt}
        \setlength\figureheight{0.7\linewidth}
        \setlength\figurewidth{1.\linewidth}
%
\begin{tikzpicture}

\begin{axis}[%
width=0.951\figurewidth,
height=\figureheight,
at={(0\figurewidth,0\figureheight)},
scale only axis,
axis on top,
xmin=0.0000000000,
xmax=20.9439510239,
xtick={\empty},
ymin=-3.1415926536,
ymax=9.4247779608,
ytick={\empty},
axis background/.style={fill=white}
]
\addplot [forget plot] graphics [xmin=-0.0102265386, xmax=20.9337244853, ymin=-12.5786424607, ymax=12.5540987681] {dist501-1.png};
\end{axis}
\end{tikzpicture}%
        \caption{$t=50$}
    \end{subfigure}
    \hfill
    }
    \caption{Time evolution of the distribution function for the bump--on--tail test case with Knudsen number $\Kn = 10^2$. Map and source term grid sizes $256^2$, sampling grid $512^2$.}
    \label{fig:bumptail_Kn1e2}
\end{figure}

\begin{figure}[p!]
\centering
{
\graphicspath{{figs/BumpOnTail_Kn1e2}}

\setlength\figureheight{0.495\linewidth}
\setlength\figurewidth{\linewidth}
\input{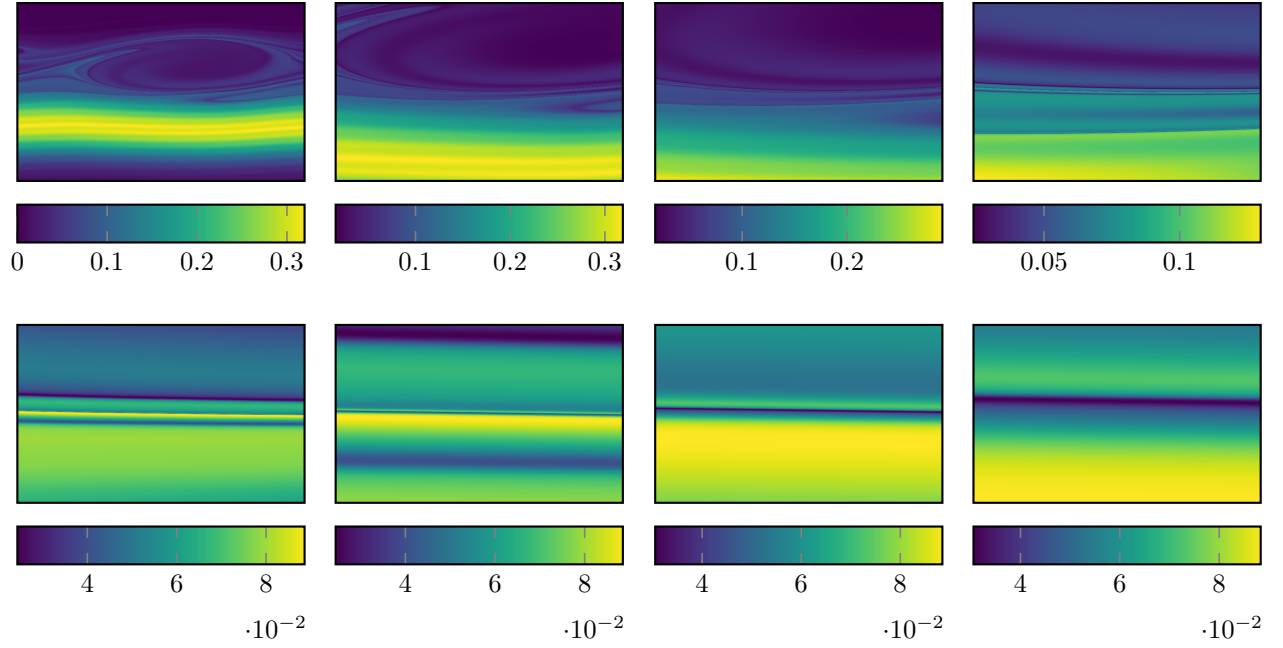}

}
\caption{Gradual zoom into the time $t=50$ density solution for the bump--on--tail instability test with Kn = $10^2$. All figures are centered on the point $(x,y) = (12.05, 2.0685)$ with box size $2h \times h$ where $h = 20, 10, 6, 2, 0.5, 0.1, 0.02, 0.005$.}
\label{fig:zoom_bumptail_Kn1e2}
\end{figure}

\begin{figure}[h]
    \centering
    {
    \captionsetup[subfigure]{justification=centering}
    \begin{subfigure}[t]{0.3\linewidth}
        \captionsetup{aboveskip=-2pt}
        \setlength\figureheight{0.6\linewidth}
        \setlength\figurewidth{\linewidth}
        \input{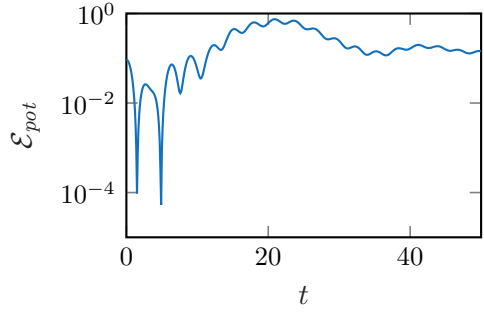}
        \caption{Potential energy}
    \end{subfigure}
    \hspace{0.2\textwidth}    
    \begin{subfigure}[t]{0.3\linewidth}
        \captionsetup{aboveskip=-2pt}
        \setlength\figureheight{0.6\linewidth}
        \setlength\figurewidth{\linewidth}
        \input{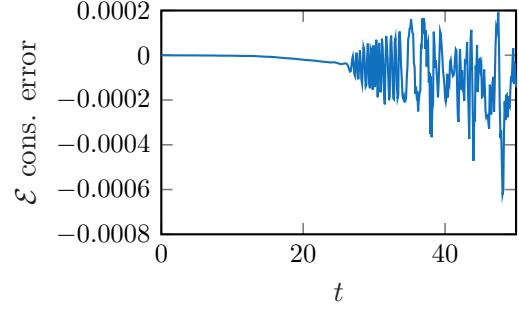}
        \caption{Total energy conservation}
    \end{subfigure}
    \hfill
    \vspace{1em}

    \begin{subfigure}[t]{0.3\linewidth}
        \captionsetup{aboveskip=-2pt}
        \setlength\figureheight{0.6\linewidth}
        \setlength\figurewidth{\linewidth}
        \input{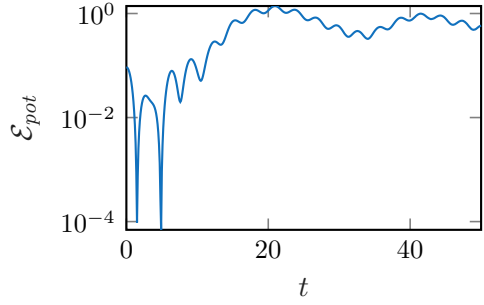}
        \caption{Potential energy}
    \end{subfigure}
    \hspace{0.2\textwidth}    
    \begin{subfigure}[t]{0.3\linewidth}
        \captionsetup{aboveskip=-2pt}
        \setlength\figureheight{0.6\linewidth}
        \setlength\figurewidth{\linewidth}
        \input{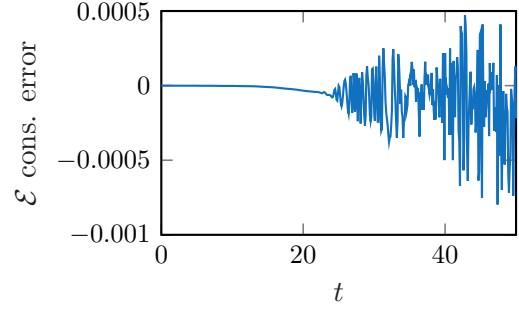}
        \caption{Total energy conservation}
    \end{subfigure}
    \hfill
    \vspace{-1em}
    }
    \caption{Potential energy evolution and total energy conservation error over time for the bump--on--tail instability tests, $\Kn = 10^2$ for the first row and $\Kn = 10^3$ for the second row.}
    \label{fig:bumptailEnergy}
\end{figure}

The bump--on--tail test cases also showcase the arbitrary resolution property of the CMM methods, even for equations with source terms. We show this by zooming into a region of the time $t=50$ solution by evaluating the map-source-term compositions \eqref{eqn:recursiveCompositionEvaluation} on a gradually smaller domain. Figures \ref{fig:zoom_bumptail_Kn1e2} and \ref{fig:zoom_bumptail_Kn1e3} show that the methods maintain arbitrary spatial resolution due to the relabeling symmetry and reduces numerical dissipation. The time evolution for the potential energy and total energy conservation in Fig.~\ref{fig:bumptailEnergy} also support the stability and conservation properties of the method.

\begin{figure}[p!]
\centering
{
\graphicspath{{figs/BumpOnTail_Kn1e3}}

\setlength\figureheight{0.495\linewidth}
\setlength\figurewidth{\linewidth}
\input{figs/BumpOnTail_Kn1e3/fz_tiled}

}
\caption{Gradual zoom into the time $t=50$ density solution for the bump--on--tail instability test with Kn = $10^3$. All figures are centered on the point $(x,y) = (12.05, 1.2291)$ with box size $2h \times h$ where $h = 20, 10, 6, 2, 0.5, 0.1, 0.02, 0.005$.}
\label{fig:zoom_bumptail_Kn1e3}
\end{figure}

\section{Conclusion}
\label{sec:conclusion}

We presented an extension of the characteristic mapping method for kinetic equations with source terms. 
For modeling collisions, we considered the local relaxation model BGK.
We proposed an original method for evaluating the Duhamel integral. 
Using the group structure, we derived a recursive formula for the time decomposition of the map and the source term integral. 
Thus the computational efficiency was enhanced by storing only sub-integrals.

As applications we considered Boltzmann and Vlasov--Poisson equations with BGK source terms for different Knudsen numbers. For time integration we proposed a diagonally implicit Runge--Kutta scheme and a related time integration approach using exponential coordinates; both methods showed third order spatial and temporal global truncation error in our convergence tests.
For benchmarking the Boltzmann--BGK solver we performed numerical simulations for the Sod shock tube problem. In- and out-flow Dirichlet boundary conditions were imposed by applying an explicit shift to the map in the flow direction.
For Vlasov--Poisson--BKG we studied the nonlinear Landau damping problem and showed results consistent with a reference pseudospectral simulation.
Simulations for a Knudsen number of $10^{-4}$ confirmed that the stiffness of the source term has been removed by the exponential time integration.
Computations of the bump--on--tail instability illustrated the outstanding fine-scale resolution capabilities of CMM using coarse grids for the map.
Moreover, the semi-group structure allowed us to zoom into the fine-scale filaments of the numerical solution.

There are different perspectives for future work. 
The extension of CMM for kinetic equations in higher dimensions in phase space is currently under way for Vlasov--Poisson, first without collisions and then with a BGK source term.
Adding more general source terms remains a challenging task, for instance more complicated non-local collisions that lead to integrals are computationally demanding and time integration might be adapted.
Considering two species and taking electrodynamic effects into account by adding Maxwell equations are also planned in prospective investigations.

\section*{Credit authorship contribution statement }
In the following, we declare the authors' contributions to this work.

{\small
\noindent
\begin{tabular}{l l}
\begin{minipage}[t]{0.18\linewidth} \textbf{X.-Y. Yin:} \end{minipage} &  \begin{minipage}[t]{0.77\linewidth} \setlength{\baselineskip}{5pt} scheme design, implementation, numerical studies, numerical analysis, visualization, writing, reviewing \& editing \end{minipage} \\
\begin{minipage}[t]{0.18\linewidth} \textbf{P. Krah:} \end{minipage} & \begin{minipage}[t]{0.77\linewidth}\setlength{\baselineskip}{5pt} initial idea, scheme design, implementation, numerical studies, writing the initial draft \& visualization  \end{minipage} \\
\begin{minipage}[t]{0.18\linewidth} \textbf{Z. Lin:} \end{minipage} &  \begin{minipage}[t]{0.77\linewidth}\setlength{\baselineskip}{5pt} implementation, numerical studies, visualization, reviewing \& editing  \end{minipage} \\
\begin{minipage}[t]{0.18\linewidth} \textbf{J.C. Nave:} \end{minipage} &  \begin{minipage}[t]{0.77\linewidth}\setlength{\baselineskip}{5pt} editing \& supervision \end{minipage} \\
\begin{minipage}[t]{0.18\linewidth} \textbf{K. Schneider:} \end{minipage} &  \begin{minipage}[t]{0.77\linewidth}\setlength{\baselineskip}{5pt} writing the initial draft, editing and supervision, funding acquisition \& project administration \end{minipage}
\end{tabular}
}

\section*{Acknowledgments}
The authors acknowledge financial support by the project ANR-20-CE46-0010, CM2E.
The authors were granted access to the HPC resources of IDRIS under the Allocation A0142A14152, project No. 2023-91664 attributed by GENCI (Grand \'Equipement National de Calcul Intensif) and of Centre de Calcul Intensif d’Aix-Marseille Universit{\'e}.
ZL and KS acknowledge the financial support from the French Federation for Magnetic Fusion Studies (FR-FCM) and the Eurofusion consortium, funded by the Euratom Research and Training Programme under Grant Agreement No. 633053. The views and opinions expressed herein do not necessarily reflect those of the European Commission. 
ZL, JCN and KS thankfully acknowledge partial financial support from the ANR-25-CE46-2339 CROKE.
JCN would like to acknowledge partial financial support from the NSERC Discovery Grant program.

\appendix
\section{Fourier pseudo-spectral reference method}
\label{sec:app2}

The CMM results are compared with a Fourier pseudo-spectral method, which serves as reference computation.  
This method evolves the particle distribution function directly on a
uniform phase-space grid, rather than evolving a characteristic map.  
The unknown is represented by its Fourier coefficients in the periodic variables
$x$ and $v$.  
On the truncated computational velocity interval $[-L_v,L_v]$,
this can be written as
\begin{equation}
  f(t,x,v)
  \approx
  \sum_{k}
  \sum_{\ell}
  \widehat{f}_{k,\ell}(t)
  e^{2\pi i k x/L_x}
  e^{i \ell \pi v/L_v}.
  \label{eq:app_spectral_expansion}
\end{equation}
Here $L_x$ is the length of the periodic spatial domain, $L_v$ is the
half-width of the truncated velocity interval, and $k$ and $\ell$ are integer
Fourier modes 
with $ 0 \le  k \le N_x  -1$ and $ 0 \le \ell \le N_v - 1$.  
The Fourier coefficients are defined by
\begin{equation}
  \widehat{f}_{k,\ell}(t)
  =
  \frac{1}{2L_xL_v}
  \int_0^{L_x}\int_{-L_v}^{L_v}
  f(t,x,v)
  e^{-2\pi i kx/L_x}
  e^{-i\ell\pi v/L_v}\d v\d x.
  \label{eq:app_spectral_coefficients}
\end{equation}
In the numerical implementation, these Fourier coefficients are computed on
the uniform collocation grid using a two-dimensional Fast Fourier Transform
(FFT).
On the velocity domain, $v$ is understood as its smooth periodic continuation,
as described in Appendix A of \cite{KrahYinNaveBergmannSchneider2024}.
Spatial derivatives are then computed by multiplication in Fourier space:
\begin{equation}
  \widehat{\partial_x f}_{k,\ell}
  =
  \frac{2\pi i k}{L_x}\,\widehat{f}_{k,\ell},
  \qquad
  \widehat{\partial_v f}_{k,\ell}
  =
  \frac{i \ell \pi}{L_v}\,\widehat{f}_{k,\ell}.
  \label{eq:app_spectral_derivatives}
\end{equation}
Products such as $v\,\partial_x f$ and $E\,\partial_v f$ are not formed by
convolutions of Fourier coefficients.  Instead, the coefficients
$\widehat{f}_{k,\ell}$ are first multiplied by the corresponding factors in
Eq.~\eqref{eq:app_spectral_derivatives}.  The inverse Fourier expansions of the
derivatives are
\begin{equation}
  \begin{aligned}
    \partial_x f(t,x,v)
    &\approx
    \sum_k\sum_\ell
    \frac{2\pi i k}{L_x}\widehat{f}_{k,\ell}(t)
    e^{2\pi i kx/L_x}e^{i\ell\pi v/L_v},
    \\
    \partial_v f(t,x,v)
    &\approx
    \sum_k\sum_\ell
    \frac{i\ell\pi}{L_v}\widehat{f}_{k,\ell}(t)
    e^{2\pi i kx/L_x}e^{i\ell\pi v/L_v}.
  \end{aligned}
  \label{eq:app_spectral_inverse_fft}
\end{equation}
The derivatives are evaluated at the physical collocation points and then
multiplied pointwise by $v$ and $E$, respectively.  The Fourier coefficients of
the products are defined by
\begin{equation}
  \begin{aligned}
    \widehat{v\,\partial_x f}_{k,\ell}(t)
    &=
    \frac{1}{2L_xL_v}
    \int_0^{L_x}\int_{-L_v}^{L_v}
    v\,\partial_x f(t,x,v)
    e^{-2\pi i kx/L_x}e^{-i\ell\pi v/L_v}\d v\d x,
    \\
    \widehat{E\,\partial_v f}_{k,\ell}(t)
    &=
    \frac{1}{2L_xL_v}
    \int_0^{L_x}\int_{-L_v}^{L_v}
    E(t,x)\,\partial_v f(t,x,v)
    e^{-2\pi i kx/L_x}e^{-i\ell\pi v/L_v}\d v\d x.
  \end{aligned}
  \label{eq:app_spectral_forward_fft}
\end{equation}
In the numerical implementation, the inverse Fourier sums are evaluated by an
inverse FFT.  The products are formed at the collocation points and transformed
back to Fourier space by a forward FFT.
This sequence of Fourier differentiation, inverse transformation, pointwise
multiplication, and forward transformation is the pseudo-spectral step, see e.g., 
\cite{Canuto2007}.

The electric field is also obtained spectrally.  The density and its spatial
Fourier coefficients are defined by
\begin{equation}
  \begin{aligned}
    \rho(t,x)
    &=
    \int_{-L_v}^{L_v} f(t,x,v)\d v,
    \\
    \widehat{\rho}_k(t)
    &=
    \frac{1}{L_x}
    \int_0^{L_x}
    \rho(t,x)e^{-2\pi i kx/L_x}\d x.
  \end{aligned}
  \label{eq:app_spectral_density}
\end{equation}
The density integral is evaluated by a uniform sum over the velocity grid, i.e., the Trapezoidal rule, and
its spatial Fourier coefficients are computed by FFT.
The electric-field relation in Eq.~\eqref{eqn:VP_BGK} is then evaluated mode by
mode.  For nonzero spatial wave numbers,
\begin{equation}
  \widehat{E}_k(t)
  =
  -\frac{i L_x}{2\pi k}\widehat{\rho}_k(t),
  \qquad
  k\neq 0,
  \qquad
  \widehat{E}_0(t)=0,
  \label{eq:app_spectral_poisson}
\end{equation}
where the zero mode is fixed by the neutrality condition.  The electric field
is recovered from
\begin{equation}
  E(t,x)
  \approx
  \sum_k \widehat{E}_k(t)e^{2\pi i kx/L_x}.
  \label{eq:app_spectral_electric_field}
\end{equation}
With these definitions, the Fourier semi-discretization is based on
\begin{equation}
  \partial_t f
  =
  -v\,\partial_x f
  -E\,\partial_v f
  +\frac{1}{\tau}\left(M[f]-f\right).
  \label{eq:app_pseudospectral_semidiscrete}
\end{equation}
At each time-integration stage, the right-hand side is evaluated at the
collocation points and transformed to Fourier space.  The standard two-thirds
truncation of the retained Fourier modes is then applied to control aliasing
errors \cite{Canuto2007}.

Because the velocity interval is finite, the pseudo-spectral reference uses the
smooth periodic continuation of $v$ described above.
This keeps the interior dynamics unchanged while avoiding a discontinuity at
$v=\pm L_v$, which would otherwise create spurious Fourier oscillations.  Time
advancement of the transport part is
performed by an explicit third-order Runge--Kutta discretization.  In the reference runs
used here, the BGK contribution is included as the source term in the
pseudo-spectral right hand side.  Thus the reference computation is a grid-based
Fourier method for the distribution itself, whereas CMM stores the transport map
and reconstructs $f$ by pullback plus accumulated source history.

\bibliography{refs2}

\end{document}